\documentclass[aps,preprint,nofootinbib]{revtex4}

\usepackage{graphics,color}
\usepackage{epsfig}
\usepackage{dcolumn}% Align table columns on decimal point
\usepackage{bm}% bold math

\begin{document}

\title{Charm Physics$-$A Field Full with Challenges and
Opportunities}
\author{Xue-Qian Li$^{1}$}
\author{Xiang Liu$^{2}$}
\author{Zheng-Tao Wei$^{1}$}
\affiliation{$^1$Department of Physics, Nankai University,
Tianjin, 300071, China\\
$^2$Centro de F\'{\i}sica Computacional, Departamento de
F\'{i}sica, Universidade de Coimbra, P-3004-516, Coimbra,
Portugal}

%             %  but any date may be explicitly specified

\begin{abstract}

In this review, we discuss some interesting issues in charm
physics which is full with puzzles and challenges. So far in the
field there exist many problems which have not obtained
satisfactory answers yet and more unexpected phenomena have been
observed at the present facilities of high energy physics. Charm
physics may become an ideal place for searching new resonances and
studying non-perturbative QCD effects, moreover probably is an
area to explore new physics beyond the Standard Model. More data
will be available at BESIII, B-factories, LHC and even future ILC
which may open a wide window to a better understanding of the
nature.

\end{abstract}

%\pacs{12.39.-x, 12.38.-t}

\maketitle

\section{Introduction}

The charm quark was a long-expected member of the quark family. It
is noted that if only $u$ quark is the intermediate fermion at the
s-channel, the cross section of the scattering $s+W^+\rightarrow
d+W^+$ would increase with the incoming energy and is unacceptable
in physics. It demands existence of another species of quarks
having the same charge as $u$ quark which serves as an additional
intermediate fermion to compensate the bad high energy behavior
for the process, i.e. retain the unitarity, this new species is
the charm quark \cite{unitarity}. Moreover, without the charm
quark the anomaly in the electro-weak model cannot be cancelled
\cite{anomaly}, so that the renormalizability of the whole theory
would be spoiled. Later by studying the $K^0-\bar K^0$ mixing,
Gaillard and Lee \cite{Gaillard} estimated that the mass of the
charm quark should be around 1.5 GeV. Thus all the urgency of
saving the beautiful theory appealed to discover this charming
"charm" quark. Then the discovery of $J/\psi$ meson and other
members of the $\psi$ family became a milestone of particle
physics \cite{Ting-Richter}. The ground state of the family are
$J/\psi$ and $\eta_c$ whose mass is about 3.1 GeV and 2.98 GeV, so
roughly it implies that $m_c\sim 1.5$ GeV which is amazingly
consistent with Gaillard and Lee's estimate. But definitely it is
not the end of story for the quark family, the discovery of bottom
quark requires existence of its partner which evaded observation
for a very long while until it was eventually found at the
TEVATRON \cite{TEVATRON} and possesses an astonishing heavy mass
about 176 GeV. Till then the three generation structure of quarks
and leptons seems complete, even though the fourth generation of
quarks and leptons is still under discussions.

We know, the $u$, $d$ and $s$ quarks reside in a triplet of the
global SU(3) quark model which successfully describes relevant
phenomenology, but there is, so far, not a special symmetry to
associate the rest three heavy quarks\footnote{There indeed is the
so-called Heavy Quark symmetry $SU_f(2)\otimes SU_s(2)$ (we will
discuss later) to connect the $b$ and $c$ quarks, but that
symmetry is a symmetry when the quark mass approaches infinity and
mainly can simplify the calculation of transition from $b$ to $c$.
It is not like the SU(3) for light flavors and is also not a
symmetry in the common sense.}. The charm quark may be a special
one in the quark family, because it is heavier than the first
three light quarks and does not belong to the regular flavor
SU(3), but stands in a weak doublet with the light strange quark.
The charm is not light at all, but also not too heavy as the
bottom and even top quarks. The intermediate mass determines the
special characteristics of hadrons which contain charm and/or
anti-charm. Recently, some researches count that the first four
quarks are different from the last two heavy ones: bottom and top
quarks, such as the top assisted technicolor model, but it is
beyond the scope of our review.

Since charm quark is indeed sufficiently heavy (will discussed
below, "heavy" means it is heavier than the binding energy scale
$\Lambda_{QCD}$), it is natural to use the potential model to
evaluate the spectra of $J/\psi$ and other family members: $\eta_c$,
$\chi_c$ and even $h_c$ etc. and their excited states, as well as
their fine-structures and decay modes. However, since it is not too
heavy, the relativistic correction to the potential which generally
consists of a Coulomb piece and a confinement one (for example the
linear potential) \cite{potential}, is more serious than that for
$\Upsilon$ family. Even though, by adjusting parameters, great
success has been achieved for the heavy charm-quarkonia (charmonia)
and the results are satisfactorily consistent with data for ground
and lower excited states. Very recently Voloshin gives an
enlightening review on charmonium where relevant topics are
discussed in details \cite{Voloshin}. However, the story is far from
its end yet. There have been many puzzles in the field, especially
as more accurate measurements are done, they do not disappear.
Moreover, several new resonances have been observed and they seem
not easy to be described by the simplest valence quark structure,
i.e. meson is composed of quark-antiquark and baryon is composed of
three quarks.

Now, let us present a rough list about the puzzles. The first one
may be the famous $\rho\pi$ puzzle where the branching ratio of
$\psi'\rightarrow\rho\pi$ is too small compared with
\cite{rhopi-exp}. Then the decay mode  $J/\psi\rightarrow\rho\pi$
is forbidden by the hadronic helicity conservation, thus its rate
must be sufficiently small, but by contraries, it is one of the
main decay channels of $J/\psi$. The sizable $D^0-\bar D^0$ mixing
which was recently observed \cite{DDmixing}, indicates that there
must exist a mechanism beyond the Standard Model (SM). There are
many newly observed resonances, which may demand interpretation.

On the theoretical aspect, great efforts have been done to look for
reasonable explanations. The first step is to make sure whether the
puzzles in the new observations are due to the flaws of our
theoretical framework or there is new physics beyond the SM. Indeed,
to understand the experimental measurements, one must calculate the
corresponding quantities, where evaluation of the hadronic matrix
elements is the key  and obstacle. Since the hadronic matrix
elements are fully governed by the non-perturbative QCD effects
which at present there is not a reliable way to deal with, all the
results we have achieved must possess certain uncertainties. To
properly estimate the errors and reliability of the results becomes
an important issue in theoretical calculations.

Below, we will discuss all the topics in separate sections. We first
recall the mechanisms which govern the weak transitions of $D$
mesons, and then focus on discussion of the rare decays because the
regular Cabibbo-favored channels have been thoroughly investigated
in both theory and experiment. Then since all the topics are related
to theoretical evaluation of the hadronic matrix elements, in next
section we review the status. In section \ref{sec4}, we concern the
final state interaction, especially focusing on the hadronic
triangle calculation and briefly discuss other schemes. In section
\ref{sec5}, we will discuss the hadronic helicity selection rule and
its violation. In section \ref{sec6} we concern the $\rho\pi$ puzzle
and in section \ref{sec7}, we show how the QCD multi-expansion
theory works very well for the pion radiation from excited states of
$\Upsilon$, but has difficulties for $\psi(nS)\to\psi(mS)+\pi\pi$
($n>m$). In section \ref{sec8}, in some details, we discuss the
newly observed resonances at the charm energy region and some of
them seem to be exotic and need a reasonable interpretation. In
section \ref{sec9}, we discuss the $D^0-\bar D^0$ mixing and related
theoretical proposals, meanwhile we also concern the possible CP
violation observable. In section \ref{sec10}, we consider the
charmed baryons and double heavy baryons $\Xi_{cc}$, where the
difference between the lifetimes of $\Lambda_c$ and $D^{\pm},\,D^0$
is especially concerned and we will also briefly discuss the charmed
pentaquark. In section \ref{sec11}, we specially discuss the diquark
structure in baryons, especially in the heavy baryons, because it is
an important issue for studying baryons. The last section is devoted
to a brief discussion.

\section{The effective Lagrangian and transition
amplitudes}\label{sec2}

Generally, this is a topic which is familiar to most of theorists
working in this field. Thus after a short introduction, we will
turn our attention to the application of the theoretical framework
for investigating rare decays of $D$ mesons and $J/\psi$.

\subsection{The decay constants of $D_s$ and $D^{0,\pm}$}

The most important parameters for the weak decays of D and $D_s$
mesons are their decay constants. Generally the decay constant of a
pseudoscalar meson is measured via its leptonic decays $P\rightarrow
l\bar\nu$ ($l=e,\ \mu,\ \tau$ as long as kinematics allows.) and the
width is written as \cite{Rosner1}
$$\Gamma(P\rightarrow l\bar\nu)={G_F^2\over
8\pi}f_P^2m_l^2M_P(1-{m_l^2\over M_P^2})|V_{qq'}|^2,$$ where
$M_P,m_l$ are the masses of the pseudoscalar meson and the lepton,
$V_{qq'}$ is the corresponding CKM entry and $q,q'$ are the valence
quarks in the pseudoscalar meson. There $f_P$ is the decay constant
which we are going to obtain. The CLEO collaboration has achieved
$$f_{D^{+}}=(222.6\pm 16.7^{+2.8}_{-3.4})~ {\rm MeV}.$$ in the decay of
$D^+\rightarrow \mu^+\nu$ \cite{CLEO-f}.

Rosner \cite{Rosner1} has obtained the average of $f_{D_s}$ as
$$f_{D_s}=(274\pm 10)~{\rm MeV}.$$ There is a discrepancy of about three standard
deviations between this result and a recent unquenched lattice
calculation \cite{LQCD}.

Very recently, the CLEO collaboration reported their new result as
\cite{CLEO-new}
$$f_{D^+}=(205.8 \pm 8.5\pm 2.5) \, \mathrm{MeV},$$ by assuming
$|V_{cd}|=|V_{us}|$. They also obtained
$$\frac{B(D^+\rightarrow \mu^+\nu)-B(D^-\rightarrow \mu^-\bar\nu)}
{B(D^+\rightarrow \mu^+\nu)+B(D^-\rightarrow \mu^-\bar\nu)}=0.08\pm
0.08.$$ which means that no CP violation was observed in the
leptonic decay. The Belle group reported \cite{Golob}
$$f_{D_s}=(275\pm 16 ({\rm stat})\pm 12 ({\rm syst})) \, \mathrm{MeV}.$$

Only when the decay constants are accurately measured, the
theoretical predictions on the hadronic transitions can be
trustworthy, so that more precise experiments are necessary. In
fact, measurement on decay constants is not easy, not only because
the so far available database on $D$, especially $D_s$ is not
large enough to guarantee a high statistics, but also reactions,
such as $P\rightarrow l\bar\nu +\gamma$ can influence extraction
of $f_P$. The BESIII will collect the largest database of $D$
mesons, so one may expect to get very accurate decay constants of
$D$ and $D_s$.

\subsection{The effective Lagrangian of weak interaction}

The charmonia $J/\psi,\ \eta_c$ and their excited states mainly
decay via strong interaction which is the OZI suppressed (will be
discussed in later sections). Instead, $D$ mesons decay via weak
interactions or electromagnetic radiation. In this framework, the
semileptonic decays are well understood, so that we concentrate
ourselves here on the non-leptonic decays. Let us briefly review
the general situation of the $D$ decays. The effective Lagrangian
for weak interaction is written as
$$L_{eff}^{|\Delta c|=1}={G_F\over\sqrt
2}\Big[V_{cs}^*V_{uq}(C_1O_1+C_2O_2)-V_{cb}^*V_{qb}\sum_{i=3}^{10}C_iO_i\Big]+h.c.,$$
where $q$ can be either $d$ or  $s$, and the first operator
$O_1=(\bar sc)_{V-A}(\bar uq)_{V-A}$ with $(\bar qq')_{V-A}\equiv
\bar q\gamma_{\mu}(1-\gamma_5)q'$ originates from the tree level
while others are induced by loops, $O_{3}$ to $O_6$ are the strong
penguin operators and the rest ($i=7$ to $10$) are due to the
$\gamma,\ Z-$penguins and box diagram \cite{He-Deshpande}.

\subsection{Rare decays of $D$ meson}

The weak decays of $D$ mesons can be categorized into Cabibbo
favored, Cabibbo suppressed and doubly suppressed modes. The first
type includes the modes such as $D^+\rightarrow \bar K^0+\pi^+$
etc. and the second one was discussed by Abbott, Sikivie and Wise
\cite{Abbott}. For the Cabibbo favored decay modes, one usually
only considers the so called spectator mechanism where the light
quark behaves as a spectator when the heavy charm quark transits
into $s$ quark plus a $u\bar d$ pair. In this picture annihilation
and $W$-exchange between charm and light anti-quark can be
neglected because of the linear momentum matching (namely, for
annihilation and $W$-exchange, a quark-anti-quark pair must be
produced from vacuum, or in other words are produced by soft
gluons, so that their linear moneta are small whereas the
quark-antiquark pair occurring directly from the effective vertex
possess large linear momenta. As they combine a quark (anti-quark)
emerged from vacuum to constitute a hadron where the two
constituents must have close linear momenta the large momentum
difference would greatly suppress the probability.). However, for
the Cabibbo-suppressed or even doubly-suppressed modes, the
annihilation and $W$-exchange mechanisms and as well as the
penguin contributions become important. In fact, there are several
channels where the spectator mechanisms do not contribute at all,
thus these modes would be ideal places to study such small effects
which may manifest some unknown mechanisms. Moreover, for the CP
violation, the contribution from the penguin and even the
electro-weak penguin would be crucially significant. We will
discuss these issues in the following sections.

Another interesting rare decay mode are those processes, where the
light quark (anti-quark) transits while the heavy charm behaves as
a spectator which only provides a color source. Such reaction
includes $D^*\rightarrow D+\gamma$, $D^*\rightarrow D\pi$ etc.
where the photon and pion can be emitted from either charm or
light flavor. We used to study a special case that a heavy baryon
containing two heavy quarks radiates a photon and transits into a
lower states with the same flavor \cite{w-Dai} in terms of the
Bether-Salpeter (B-S) equation. It was indicated that the
branching ratios of such decays are very small and hard to measure
at the present luminosity, however, for BESIII and LHCb, the
situation may be greatly improved.

Recently, Li and Yang \cite{Li-Yang} calculated the branching
ratios of $D^+\rightarrow D^0+e^++\nu$, $D_s^+\rightarrow
D^0+e^++\nu$, $D_s^+\rightarrow D^++e^++e^-$ in SM, however, their
results indicate that only the branching ratio of
$D_s^+\rightarrow D^0+e^++\nu$ could reach $10^{-8}$. According to
the sensitivity of BESIII, B-factories, Super-B and LHCb, it might
be observed at Super-B and LHCb, but not at others. On other side
the observation may offer a probe for testing the working
mechanisms which govern the behaviors of the light flavors in
hadrons which are usually treated as passive spectators in most of
reactions, as aforementioned.

An interesting discovery  draws attention of theorists, it is the
observation of baryonic decay $D_s\rightarrow p\bar n$
\cite{exp-Ds-pn}, which can only occur through the
$W$-annihilation topology, so it was supposed to be very
suppressed as aforementioned for the meson case \cite{Cheng1}.
Chen et al. \cite{Cheng2} indicated that the short-distance
contribution can only make the branching ratio as large as
$10^{-6}$ which is much smaller than the data. Thus they suggested
that the long-distance contribution via the FSI can enhance this
value, so that they claimed that it is a dynamical enhancement of
the $W$-annihilation topology in $D_s$ decays. It is worth further
studies indeed.

\subsection{Weak decays of $J/\psi$}

This is another type of rare decays which may provide us with some
information about the structure of $J/\psi$. Generally $J/\psi$
would decay via strong  or electromagnetic interactions. The strong
decay is realized via a process where the constituents $c$ and $\bar
c$ annihilate into three gluons which eventually fragment into
hadrons, it is an OZI-suppressed reaction and that is also why
$J/\psi$ is a narrow resonance and evaded observation before 1974.
It is believed that the decay width is proportional to the
wavefunction of $J/\psi$ at origin which can be easily obtained by
measuring its leptonic decay width. However (see below), a violation
of the hadronic helicity selection rule indicates that such a
picture may be not completely correct, therefore to investigate the
structure of $J/\psi$ (if it has a hybrid component etc.), study on
the weak decay of $J/\psi$ might be very helpful. We calculated the
branching ratio of the semi-leptonic decay $J/\psi\rightarrow
D_s^{(*)}+e^++\nu$ \cite{y.Wang1} in the QCD sum rules and obtained
it to be of order of $10^{-10}$. Then with the gained parameters, we
extended our calculation to the non-leptonic weak decays of
$J/\psi$. The results show that the branching ratio of inclusive
weak decays can reach order of $10^{-8}$, and a special channel
$J/\psi\rightarrow D_s^{(*)}+\rho$ \cite{y.Wang2} has a larger
branching ratio of about $5.3\times 10^{-9}$ which might be measured
by BESIII, B-factories, Super-B and LHCb, as we wish.

In fact, all such rare decays may be important for better
understanding of the structure of $J/\psi$ and the governing
dynamical mechanisms, even though accurate measurements on them are
extremely difficult. We lay our hope on the very large database of
the facilities which will be available soon.

\section{Hadronic matrix elements}\label{sec3}

This is probably the most difficult problem in hadron physics which
almost covers the whole field of high energy physics, and is
definitely confronted by anybody. The reason is that hadronization
occurs at the energy scale below $\Lambda_{QCD}$ where
non-perturbative QCD effects dominate and so far there is no an
effective way to accurately evaluate the effects yet. Much efforts
have been made to handle the problem. The simplest way is using the
naive factorization where a hadron, generally a meson, is emitted
and can be factorized out from the hadronic transition of one hadron
(meson or baryon) to another one. Then the rest transition amplitude
can be parametrized by a few form factors which are obtained by
fitting data \cite{Stech}. The transition matrix element can be
analytically decomposed into a few terms according to the Lorentz
structure and the parity conservation because hadronization is a
process where only strong interaction applies. The advantage of this
method is that it is simple and since the form factors are fixed by
fitting data, they can be extensively applied to study the weak
decay rates. It is simple and consistent with data within a rather
wide range, however, there are obvious shortcomings. First the
factorization is not always legitimate, as Buras et al. pointed out
\cite{Buras}, the matrix element of operator
$$\langle M_1M_2|C_1\bar q_{1i}\gamma_{\mu}(1-\gamma_5)q_{2i}\bar
q_{3j}\gamma^{\mu}(1-\gamma_5)q_{4j}+C_2\bar
q_{1i}\gamma_{\mu}(1-\gamma_5)q_{2j}\bar
q_{3j}\gamma^{\mu}(1-\gamma_5)q_{4i}|M\rangle,$$ cannot simply
written as the factorized form because a term proportional to
$$\lambda^a_{ij}\lambda^a_{lm}$$
would appear and phenomenologically it causes an effective $N_c$
and the coefficients are deformed as
$$C_1+C_2/N_c^{eff}\;\;\;{\rm or}\;\;\;C_1+C_2/N_c^{eff}.$$
where $N_c^{eff}$ is no longer 3 \cite{Cheng-1}.

Moreover, such factorization is based on the spectator mechanism
where the transition occurs at the heavier quark leg and another
light component would play a role of a spectator. In this way, the
annihilation and $W$-exchange sub-processes are not properly
included. On other aspect, as we know, the annihilation and
$W$-exchange sub-processes might be important, when dealing with
the inclusive processes, especially for evaluating the lifetimes
of $D$ mesons \cite{Georgi-Carone}.

Thanks to the heavy quark effective theory (HQET) where an extra
symmetry $SU_f(2)\times SU_s(2)$ is considered, one can reasonably
evaluate the transition between two heavy mesons containing $b$ or
$c$ quarks. It has already becomes a criterion for testifying
validity of any theoretical calculations where hadronic matrix
elements are evaluated, as if the heavy quark limit is taken, the
results must be qualitatively consistent with that obtained by the
HQET. Moreover, Georgi generalized this scenario for dealing with
transition between two heavy baryons, each of which contains two
heavy quarks in terms of the superflavor symmetry
\cite{Guo-Jin-Li}. On another side, charm is not heavy enough to
be treated as a real heavy quark, and the $1/m_c$ corrections may
be important, even for heavier b-quark, the $1/m_b$ corrections
are not negligible in  practical computations \cite{1/mb}.
Therefore, just as the results under large-$N_c$ limit in the
$1/N_c$ expansion theory only possess qualitative meaning, the
obtained values under heavy quark limit (i.e. let $m_Q\to\infty$)
correspond to the leading order, corrections must be accounted
while comparing with more accurate experimental data. There are
many works to consider how to properly evaluate the $1/m_Q$
corrections \cite{1/mb}.

In the recent years, some effective theories have been developed to
justify the factorization, especially in $B$ physics. The
perturbative QCD (pQCD) is based on the factorization theorem. It
states that for a process with large momentum transfer, the physical
amplitude can be factorized as a product $\phi\otimes H$ where $H$
is the factor corresponding to the quark-level hard process
amplitude which can be calculated in perturbation theory order by
order and $\phi$ stands for the soft part. The later one is not
calculable in the perturbative way. In general the product is
related to a convolution integration over the wave functions of the
initial and final hadrons. In the literatures, there are two
different versions of pQCD to deal with the factorization, one is
the familiar collinear factorization and another is the so-called
$k_{T}$ factorization. The crucial difference between the two
schemes may be that $k_T$ factorization is possible to treat the
problem of endpoint divergence. The details of the two schemes can
be found in relevant literature \cite{Li-H-n}. The soft-collinear
effective theory (SCET) is a recently developed effective field
theory to simplify the processes containing light energetic hadrons
\cite{scet,Lu}. The great development is that the proof of the
factorization theorem can be performed at the operator level in the
SCET which is much simpler than the diagrammatic analysis in pQCD.
However, on the other hand, except very few processes, such as the
pion transition form factor, have been proved to be factorizable,
most exclusive processes cannot be rigorously proved to be
factorizable. Moreover, the factorization proofs are usually limited
to the leading order and therefore the application factorization
theorem is still an assumption. The factorization may be applicable
for the decays of bottomed mesons and baryons, but for charmed
mesons $D$ and baryons, it is indeed questionable. In $D$ meson
decays, the energy of final light meson is at the order of
$\Lambda_{\rm QCD}$ which is not high enough to perform a
perturbative analysis. Another question arises from the substantial
corrections in power of $\Lambda_{\rm QCD}/m_c$. All these facts
make it difficult to apply pQCD or SCET into charm decays. In
another approach named as the Transverse Momentum Distribution
(TMD), where factorization is performed by taking transverse momenta
of partons into account. For example in the reaction
$\pi^0+\gamma^*\to \gamma$ the form factor can be written as a
convolution integral
$$F(Q^2)\sim \phi\otimes S\otimes H(1+\mathcal{O}(Q^{-2}). $$
where $\phi$ is the light-cone wavefunction of the pion, $S$ is an
additional soft factor and $H$ corresponds to the hard scattering
part \cite{Ma}.

A very recent work by several authors \cite{Ma1} indicates that for
the $k_T$ factorization scheme, at loop-level  an extra term which
is related to the so-called light-cone divergence appears when a
unitary gauge is employed in the calculation and disappears in the
Feynman gauge. This explicitly manifests that the the $k_T$
factorization at loop-level is not gauge-independent, so that is
violated. This statement is still in dispute. Even though the $k_T$
factorization cannot be a strict theory according to the field
theory, it can definitely treated as a successful phenomenological
model and is applied to calculate the transition amplitudes where
heavy hadrons are involved. Therefore generally we can trust the
theoretical results which are obtained based on the $k_T$
factorization. Very, very recently, H. Li argues that the work
\cite{Ma1} might make some calculation mistake and he presented a
result which is free of the light-cone singularity, so the gauge
invariance of the $k_T$ factorization is kept \cite{H-Li-new}. Since
it is a serious dispute, we will follow the further development in
the interesting regime.

Besides these methods which may stem from the quantum field
theory, there are some traditional methods which have been widely
applied to calculate the hadronic matrix elements including for
example: the harmonic oscillator model \cite{harmonic}, the
constituent quark model \cite{Kanki}, the constitute quark meson
model (CQM) \cite{Ebert-feldmann,Polosa}, light front quark model
\cite{HW}, color-singlet model \cite{singlet}, color-octet, and
color evaporation \cite{evaporation}, especially the
non-relativistic QCD (NRQCD) where an expansion in powers of
velocity of the heavy quark $v$ is naturally made \cite{NRQCD}.
Besides these phenomenological models whose parameters must be
fixed by fitting data, the theoretical framework QCD sum rules
\cite{SVZ} is based on quantum field theory where only the
perturbative vacuum is replaced by the physical vacuum. Because of
the properties of the vacuum, a series of condensates of
quark-pair, gluons and quark-gluon etc. are introduced to describe
the non-perturbative QCD effects. It has achieved great success in
phenomenology. However, on the other side, it is an extrapolation
from the region where perturbative QCD works reliably
\cite{Shifman}.  In the expansion only the operators with lower
dimensions are retained and moreover, to extract physical results
a reasonable plateau is required, where the threshold values are
determined, thus  an error of about 15\% is unavoidable.

Quite amount of phenomenological models have also been employed to
calculate the hadronic matrix elements whose energy scale is
$\Lambda_{QCD}$. As well known, the hadronization is fully
determined by the non-perturbative QCD effects and so far there is
no a reliable way to evaluate them based on quantum field theory
or any other first principles. One of the goals of our research in
fact, are to determine the mechanism, which governs the reaction,
generally the fundamental theory is the standard model (SM) which
at present no one doubts due to its remarkable success, thus we
can reliably (if ignore contributions from new physics beyond the
SM) determine the hard factor  due to the asymptotic freedom of
QCD. To extract important information, such as determining the CKM
matrix elements and checking the unitary triangle, exploring CP
violation, one indeed needs to have a more accurate estimation of
the hadronic matrix elements, otherwise the physical picture would
be contaminated by the inaccuracy. Therefore, to understand the
physical world, reliable estimate on the non-perturbative effects
are absolutely necessary and all the efforts along the line are
worthwhile.

Moreover, one can also use the MIT bag model with taking into
account the recoil effects \cite{MIT-bag}, the chiral bag model
and even the flux-tube model \cite{flux-tube} to describe the
wavefunctions of the initial and final hadron states when carry
out the calculation of the transition matrix elements.

Indeed all the models have their own reasonability and advantage,
but there are obvious flaws and un-reasonability, and because they
are not coming from a basic principle, one can never expect that
they can be perfect. Therefore on one side, even though the models
are not perfect, they have applicability and if they are properly
applied, reasonable results should be reached.

As a conclusion, estimation of the hadronic matrix elements is
crucially important, but so far, there does not exist a way to
fulfil the job yet and one can only apply the available models to
estimate them with certain reliability.

\section{Final state interaction}\label{sec4}

Besides the estimate on the hadronic matrix elements which are
directly related to the transition, there are secondary reactions,
namely the final state interactions (FSIs) which are also very
significant for charm-hadron decays. Such processes are due to
strong interaction and occur at hadron level, thus also cannot be
derived by perturbative QCD.  Fortunately, one can use the chiral
Lagrangian to evaluate the long-distance effects and we will discuss
this issue in this section. Some phenomenological models had been
suggested to estimate the FSIs in $D$ meson decays:
one-particle-exchange model \cite{OPE} and the Regge pole model
\cite{Dai-YS}.

The final state interaction (FSI) in the charm-tau energy region
is very important \cite{Cheng-H.Y,Li-Zou}. According to the
concept, the FSI can be categorized into the quark-level and
hadron level FSI processes. The quark level FSI process refers to
the quark interference due to the identical fermion statistics. It
was noticed by Stech et al. long time ago to explain the lifetime
difference of $D^0$ and $D^{\pm}$ and we will come to this subject
in later sections. Now let us concentrate on the second category
of FSI i.e. that at hadron level. In those processes, the initial
hadron first decay into intermediate hadrons (usually two hadrons)
and the two hadrons would re-scatter into the final states. Since
the re-scattering occurs via strong interaction, the isospin must
be conserved.

The re-scattering occurs at hadron level and both of the hadrons are
in color-singlet, thus the interaction between the hadrons cannot be
described by one or even a few gluon-exchange and it makes the whole
calculation more tricky and uncertain. The responsible effective
theory in this field should be the chiral Lagrangian. However all
the coefficients in the lagrangian cannot be obtained from an
underlying theory, such as QCD at present, and can only be fixed by
fitting data. this brings up very serious problem and uncertainties
for the theoretical evaluation.

To evaluate the re-scattering effects, some authors suggested to
calculate the absorptive part of the triangle diagrams whose
internal legs are intermediate hadrons and external lines correspond
to the initial hadron and the final daughter hadrons
\cite{Anisowich}. In fact the absorptive part of the triangle
corresponds to the real Final State re-scattering because the two
intermediate hadrons which are directly coming from the initial
decaying hadron are on their mass shells. The hadron(s) exchanged
between the two intermediate hadrons at t-channel not only needs to
possess proper quantum numbers, but also has to conserve
energy-momentum, so that it is obviously off-shell. As at the
triangle apexes we apply the effective interaction vertices  which
are extracted from the chiral Lagrangian, one needs to introduce
phenomenological form factors to compensate the off-shell effects.
Usually there are various types of the form factors which are widely
adopted in literature, the simplest one is the pole
form\footnote{The usually adopted form of the form factor is
$({\Lambda^2-m^2\over \Lambda^2-q^2})^n$ where $n=1$ is the monopole
form, $n=2$ is the dipole form and as $n>2$, it is a multi-pole form
which is seldom selected in literature. Besides, there are
exponential and other forms.},
$${\Lambda^2-m^2\over \Lambda^2-q^2},$$
where $q$ and $m$ are the momentum and mass of the t-channel
exchanged hadron and $\Lambda$ is a phenomenological parameter whose
value is believed to be close to 1 GeV. It is noted that the form
factor can also paly the role of the cut-off in the Pauli-Villas
renormalization scheme, so that in the calculations, no ultraviolet
and infrared divergences bother us. On other aspect, this also leads
to parameter-dependence and makes theoretical predictions uncertain.
Therefore, this calculation can only tell us the order of magnitude
for the concerned reaction unless we can use some data as inputs to
fix the model parameters.

The importance of FSI can be understood as one studies the decays
of $D^0\rightarrow K^0\bar K^0$ and $D^0\rightarrow K^+K^-$
\cite{Dai-YS}, the former process can only occur via $W$-exchange
diagram which is very suppressed according to the general
analysis, moreover, due to a CKM cancelation, this reaction should
be proportional to an SU(3) violation, so that should be very
small, in comparison, the later one is a Cabibbo favored external
emission process and should be overwhelmingly larger than the
former one. However the data show that $B(D^0\rightarrow
K^+K^-)=(3.84\pm 0.3)\times 10^{-3}$ and $B(D^0\rightarrow
2K_S^0)\sim (3.7\pm 0.7)\times 10^{-4},$ which implies
$B(D^0\rightarrow K^0\bar K^0)$ is comparable with
$B(D^0\rightarrow K^+K^-)$. This can be easily realized via a
re-scattering $K^+K^-\rightarrow K^0\bar K^0$. In our work
\cite{Dai-YS}, we showed that it can be realized with the data
measured in experiment on $KK$ scattering as inputs. This simple
example confirms the importance of FSI in charm physics.

Moreover, the final state interaction provides a strong phase which
may lead to CP violation. As well known the direct CP violation in
decays is
$$\Gamma(A\rightarrow B)-\Gamma(\bar A\rightarrow B)\sim
\sin(\alpha_1-\alpha_2)\sin(\phi_1-\phi_2).$$ where $B$ may be a
CP eigenstate and $\alpha$, $\phi$ are strong phase and weak phase
respectively. It is obvious that there at least exist two
independent channels which have different strong and weak phases,
otherwise the direct CP asymmetry is zero. Due to the final
products may originate from another weak process and therefore the
reaction can have different weak phase from the direct transition
and the FSI can provide a strong phase, i.e. the phase shift in
the language of scattering, thus their interference can result in
a non-zero CP asymmetry \cite{Li-Zou}.

To calculate effects of the final state interaction, there are two
possible ways to adopt, i.e. the Regge-pole model and the hadronic
loop. It was discussed that the Regge model may apply in higher
energy region whereas the hadronic loop method is more suitable in
lower energy region. To certain accuracy they are consistent.

Generally, as many authors discussed, the two intermediate hadrons
are real on-shell particles, therefore, so that one only needs to
calculate the absorptive (i.e imaginary) part of the triangle
diagrams. In fact, it is the real final state interaction by the
common sense. However, as Suzuki \cite{Suzuki} pointed out, the
dispersive part of the triangle can also paly a role to influence
the transition amplitude. By calculating the dispersive part of
the triangle, we evaluate the branching ratio of
$J/\psi\rightarrow PV$ where $P$ and $V$ stand for pseudoscalar
and vector mesons \cite{Liu-Li-Zeng}. This is related to the
famous $\rho\pi$ puzzle which will be discussed in later sections.
Our strategy is that we calculate the dispersive part of the
triangle by keeping the parameter $\Lambda$ as a free parameter to
be determined, and by assuming the SU(3) symmetry we set the
direct transition rate as another free parameter, then by fitting
two special channels ($J/\psi\rightarrow \rho^0\pi^0,\; K^{+*}\bar
K^-$ which are more accurately measured) we obtain the two
parameters. With them we calculate the branching ratios of other
channels and find that the results which are listed in the table
below, are well consistent with data.

\begin{table}[htb]
\tiny
\begin{center}
\begin{tabular}{c||cccccccccc} \hline
Decay
mode&$\rho^{0}\pi^{0}$&$K^{*+}K^{-}+c.c.$&$\phi\eta$&$\phi\eta'$
&$\omega\eta$&$\omega\eta'$\\\hline\hline
BR$\times10^{-3}$(Experiment)\cite{databook}&$4.2\pm0.5$&$5.0\pm0.4$
&$0.65\pm0.07$&$0.33\pm0.04$
&$1.58\pm0.16$&$0.167\pm0.025$\\\hline $\mathcal{G}^{PV} (10^{-3}$
GeV$^{-1})$&$2.08\pm0.25$&$1.65\pm0.26$&$0.89\pm0.096$&$0.71\pm0.086$
&$1.27\pm0.13$&$0.46\pm0.069$\\\hline
$\mathcal{G}^{PV}_{H} (10^{-3}
$GeV$^{-1})$(Theory)&6.44&6.01&3.47&5.01&5.33&3.97\\\hline
$\mathcal{G}^{PV}
(10^{-3}$GeV$^{-1})$(Theory)&2.08(fitting)&1.65(fitting)&0.93&0.61&0.93&0.43\\\hline
BR$\times10^{-3}$(Theory)&4.2(fitting)&5.0(fitting)&0.71&0.25&0.84&0.15\\\hline
\end{tabular}
\end{center}
\caption{The first two modes are well measured and  the
theoretical model parameters are obtained by fitting them
\cite{Liu-Li-Zeng}.\label{pv-table}}
\end{table}
\vspace{0.6cm}

It is interesting to note that usually only at lower energy region,
the FSI effects are more significant, but sometimes, the small
effects may be also non-negligible. This is the case for studying CP
violation. As indicated, the direct CP violation is induced by an
interference between at least two channels which have different weak
and strong phases. Even though the FSI in certain cases is much
smaller than the main contribution which usually comes from the tree
level and does not possess a strong phase, FSI then provides a
non-zero strong phase and its weak phase may be completely different
from the tree level one, thus an interference between it and the
tree amplitude would result in a non-zero CP violation. We
calculated such a possibility for the $B_c$ decay \cite{Liu-Bc} and
conclude that in the case the pQCD calculation is still valid and
the FSI effect only raises a minor contribution to the total decay
amplitude, but its strong phase is not zero, so may help to build up
an observable CP violation effect as it interferes with the tree
contribution which is calculated in pQCD.

This indicates that the FSI is important, but as aforementioned, the
large uncertainties of the whole scenario and parameters involved in
the calculations make the theoretical prediction not accurate, and
all these are worth further studies. To make the theory in a better
shape, we need more information from experiment.

\section{Hadronic Helicity suppression in $J/\psi$ decays}\label{sec5}

It is generally believed that the narrowness of $J/\psi$ is due to
the Okubo-Zweig-Iizuka (OZI) rule \cite{OZI}. Violation of the OZI
rule would give rise to a non-zero transition rate. The main
two-body decays of $J/\psi$ can be categorized as
$J/\psi\rightarrow PP'$, $J/\psi\rightarrow PV$ and
$J/\psi\rightarrow VV'$ where $P$, $V$ stand for pseudoscalar and
vector mesons. Besides, definitely there are other modes, for
example, the scalar and axial vector final states, and as well,
there is also possibility to decay into two baryons such as $p\bar
p$, $\Lambda\bar\Lambda$ etc. but the corresponding branching
ratios of such channels are rather small \cite{databook} and the
result is well understood in our theory. To get a better
understanding of the OZI rule, people turn to study the radiative
decays of orthoquarkonia \cite{Korner} where only a hadronic
transition matrix element is needed. With certain approximations
they obtained numerical results which were roughly consistent with
data. The process involves a five-point Green's function and the
Feynman integration is complicated. Thanks to the developments of
the calculating techniques, we carried out a full calculation
\cite{Tong-Li} and the results are qualitatively consistent with
that obtained by the authors of Ref. \cite{Korner} and the data
available at that moment. The success indicates that our knowledge
on the OZI rule might be right with a tolerable error. It is also
noticed that in the radiative decays, the hadronic effects of
$J/\psi$ are included in its wavefunction at origin which is
obtained by fitting data of the leptonic decays of $J/\psi$ and it
must be accurate enough.

To further testify the OZI rule, we calculate the process of
$J/\psi\rightarrow \pi\pi$ which is an isospin violating one.
Usually it is supposed that this process is induced by the
electromagnetic interaction, i.e. via $J/\psi\rightarrow
\gamma^{\star}\rightarrow \pi\pi$. However, as well known, there
are two sources for isospin violation, one is the electromagnetic
interaction and another is the mass splitting of $u$ and $d$
quarks. Considering the mass difference of $u$ and $d$ quarks, we
calculate the OZI process $J/\psi\rightarrow ggg\rightarrow
\pi\pi$ \cite{Li-Tong} and got its amplitude which is comparable
with the contribution from electromagnetic interaction. Concretely
we find the OZI amplitude is proportional to
$(m_u-m_d)/M_{J/\psi}$ which clearly manifest the isospin
violation. We should will carry out similar calculation of
$J/\psi\rightarrow \gamma^*\rightarrow \pi\pi$ in the same
scenario and compare it with the OZI process in our later work.
Then we continue to calculate the branching ratio of
$J/\psi\rightarrow \rho\pi$ which is supposed to be fully
dominated by the OZI process $J/\psi\rightarrow ggg\rightarrow
\rho\pi$. We obtained a branching ratio which is one order smaller
than the data More discussion will be presented in the next
section).

In fact, it is not a surprise, because a long time ago, Brodsky and
Lepage \cite{Brodsky} indicated that as the vector-gluon coupling
conserves the quark helicities, the total hadronic helicity is
conserved and can only be violated at order of $m/Q$ or higher,
where $m$ is the light quark mass in the final hadrons and $Q$ is
the transferred momentum scale. Our numerical results are consistent
with this rule, i.e. the process is suppressed by the hadronic
helicity conservation and we explicitly show that the amplitude of
the transition is proportional to $(m_u+m_d)/M_{J/\psi}$ which
confirms the observation of Lepage et al. However, the allegation
sharply contradicts to the data where the branching ratio of
$J/\psi\to\rho\pi$ is $(1.69\pm 0.15)\%$ \cite{databook} and is one
of the dominant hadronic modes. To compromise this obvious
discrepancy, Branbilla \cite{Brodsky} suggested that either $J/\psi$
contains other constituents, for example is a hybrid, or there is
some unknown mechanism which violates the hadronic helicity
conservation and results in a larger transition amplitude. However
both of the interpretations would receive very crucial challenges
because for a long time, people believe that $J/\psi$ is composed of
charm and anti-charm quarks and almost overwhelming works are based
on this picture. The second one also does not seem to work, because
if it were true, the mechanism would exist in other channels and
influence all theoretical predictions.

Associating with the phenomenon where $\psi'\to\rho\pi$ is
peculiarly suppressed, we are inclined to the first interpretation,
namely $J/\psi$ is not a pure $c-\bar c$ bound state, but $\psi'$
is. We will give more discussions in  the later section about the
$\rho\pi$ puzzle. On other aspect we used to follow Suzuki and
consider the long-distance contribution to $J/\psi\rightarrow
\rho\pi$ and our results show that it is possible to explain data,
but the answer is still not satisfactory yet (see later section for
more discussions).We used to follow Suzuki and consider the
long-distance contribution to $J/\psi\rightarrow \rho\pi$ and our
results show that it is possible to explain data, but the answer is
still not satisfactory yet (see later section for more discussions).

\section{The $\rho\pi$ puzzle}\label{sec6}

As widely discussed, the puzzle has been raised for a long while. In
the regular theoretical framework, there should be a relation
$$R={BR(\psi'\rightarrow ggg)\over BR(J/\psi\rightarrow
ggg)}={\Gamma(\psi'\rightarrow e^+e^-)\over \Gamma(J/\psi\rightarrow
e^+e^-)}\cdot {\Gamma_t(J/\psi)\over \Gamma_t(\psi')},$$ where
$\Gamma_t$ is the total width.   This ratio comes from the fact that
if both $J/\psi$ and $\psi'$ are $c-\bar c$ bound states, in the
hadronic decays, $c$ and $\bar c$ annihilate into three gluons which
then convert into hadrons, whereas in the leptonic decays, $c$ and
$\bar c$ annihilate into a virtual photon which turns into a
lepton-pair. In this picture, the amplitudes of the hadronic decay
which occurs via a three-gluon intermediate state, and the leptonic
decay which occurs via a virtual photon intermediate state are
proportional to the wavefunction at origin $\psi(0)$. If everything
worked well, the ratio should be close to 12$\sim 14$\%, which is
called as the 14\% rule (now, it is sometimes called as 12\% rule,
anyhow it is a sizable number in contrast to the data.). However the
data tell us that this ratio is much smaller than this value.

Some theoretical interpretations have been proposed. Rosner et al.
\cite{Rosner} suggested that the quantum number of the observed
$\psi'$ may be not a pure 2S state which is the first radial
excited state of the $c\bar c$ system, but a mixture of 2S and 1D
states. The amplitudes are instead
$$\langle\rho\pi|\psi'\rangle=\langle\rho\pi|2^3S_1\rangle\cos\phi-
\langle\rho\pi|1^3D_1\rangle\sin\phi\sim
0,$$
$$\langle\rho\pi|\psi"\rangle=\langle\rho\pi|2^3S_1\rangle\sin\phi+
\langle\rho\pi|1^3D_1\rangle\cos\phi\sim
\langle\rho\pi|2^3S_1\rangle/\sin\phi,$$ where $\phi$ is fixed as
$-27^{\circ}$ or $12^{\circ}$ by fitting data. By the destructive
interference between the contributions of the two components to the
amplitude of $\psi'\rightarrow \rho\pi$, the smallness is explained.
Suzuki \cite{Suzuki} alternatively suggested that the relative phase
between the one-photon and gluonic decay amplitudes or hadronic
excess in $\psi'$ decay may result in the small branching ratio. The
final state interactions may also give a reasonable explanation
\cite{Zou-11}. The first proposal can be tested in the decays of
$\psi"\rightarrow \rho\pi$ which has not been well measured yet. In
Ref. \cite{Suzuki}, the author suggested that the one-photon
amplitude is sizable and it can be tested in some other modes, for
example $\psi\rightarrow \pi\pi$ if the process is dominated by the
electromagnetic interaction. The hadronic excess can also receive
tests in the decays of other higher excited states of $\psi-$ family
and even $\Upsilon-$ family. The final state interaction may play an
important role in $D$ and $B$ decays, and also in decays of $\psi$
mesons as suggested in literature. The difficulties are how to
properly evaluate such effects. The final state interaction process
is induced by strong interaction at lower energy region, thus it is
governed by the non-perturbative QCD which is not fully understood
in the present theoretical framework yet. People need to invoke some
phenomenological models to carry out the calculations. We will give
a more detailed discussion on estimation of final state interaction,
here we only use our results to discuss the puzzle.

In our work \cite{Liu-Li-Zeng}, we simultaneously consider the FSI
and the direct decay of $J/\psi$ into a vector and a pseudoscalar
mesons and conclude that both of them contribute to the widths and
their interference should be destructive to explain data. This
observation indicates that even though the OZI-forbidden process
is sizable, it cannot be consistent with data. The result implies
that the hadronic helicity conservation indeed greatly suppresses
the process of $J/\psi\to \rho\pi$ and as the data  demand an
explanation, one should consider what is the origin of the
problem.

In a straightforward calculation based on the SM, we estimate the
decay width of the OZI forbidden process $J/\psi\to\rho\pi$
\cite{Li-Tong}, and find that the width is indeed proportional to
$(m_q/m_{J/\psi})^2$ which is coming from the hadronic helicity
suppression factor. Numerically the branching ratio of
$J/\psi\to\rho\pi$ should be smaller than 0.1\%. The same
situation appears for $\psi'\to\rho\pi$. It was qualitatively
discussed by Brodsky et al. \cite{Brodsky}. As aforementioned in
last section, to testify the calculation, we recalculate the
subprocess $J/\psi\to 3g\to \pi\pi$, which is an isospin violating
reaction and usually is supposed to be dominated by the subprocess
$J/\psi\to\gamma\pi\pi$ because the EM interaction violates
isospin as well known. Our result indicates that in the OZI
forbidden subprocess the transition amplitude is proportional to
$(m_u-m_d)/m_{J/\psi}$, i.e. the mass difference results in the
isospin violation instead. Our numerical result is of the same
order as the contribution from $J/\psi\to\gamma\pi\pi$. All the
results are consistent with our physics picture and qualitatively
reasonable. Therefore we can trust our calculations for the
process $J/\psi\to\rho\pi$. Our numerical results are listed in
Table \ref{aa}.
\begin{table}[h]
\caption{Decay widths ($\Gamma$) of $J/\psi\to \pi^+ \rho^- +
\pi^- \rho^+$ based on the three distribution functions, $\phi_1$,
$\phi_2$ and $\phi_3$, respectively \cite{Li-Tong}.  \label{aa}}
\begin{center}
\begin{tabular}{c|c|c|c|c|c}
  \hline\hline
  $m_u$(MeV) & $m_d$(MeV) & $\Gamma(\phi_1)$(MeV) & $\Gamma(\phi_2)$(MeV)
  & $\Gamma(\phi_3)$(MeV) & exp(MeV) \\
  \hline\hline
  2 & 2 & $1.04\times 10^{-4}$ & $7.21\times 10^{-5}$ & $5.11\times 10^{-4}$ &  \\
  3 & 3 & $2.36\times 10^{-4}$ & $1.6\times 10^{-4}$ & $1.17\times 10^{-3}$ &  \\
  4 & 4 & $4.12\times 10^{-4}$ & $2.9\times 10^{-4}$ & $2.08\times 10^{-3}$
  &  $(1.06\pm 0.08)\times 10^{-3}$\\
  5 & 5 & $6.69\times 10^{-4}$ & $4.54\times 10^{-4}$ & $3.38\times 10^{-3}$ &  \\
  6 & 6 & $9.75\times 10^{-4}$ & $6.68\times 10^{-4}$ & $4.88\times 10^{-3}$ &  \\
  \hline\hline
\end{tabular}
\end{center}
\end{table}

As indicated in \cite{Brodsky}, the structure of $J/\psi$ may be not
a pure $c\bar c$ charmonium, but consists of other components, such
as hybrids $c\bar cg$, $c\bar cq\bar q$ and etc.

To understand the smallness of the ratio $R$, one can expect that
either there is a problem with $\psi'$ as Rosner et al. do, or
something obscure in $J/\psi$ structure as Brodsky and many others
indicated. Our above numerical results show that even though the
hadronic selection rule works in the cases of $J/\psi\to \rho\pi $
and $\psi'\rightarrow \rho\pi$, the suppression is not too serious
and the theoretical prediction is only one order smaller than the
data.

Therefore a tentative conclusion may be drawn that the $\rho\pi$
puzzle may be not due to the mixing structures of $\psi'$ and
$\psi"$, but neither to an anormal structure of $J/\psi$ itself. It
seems that both of the proposals cannot independently explain the
"puzzle", more complicated mechanisms may be needed.

This is a great challenge to our understanding because the $c\bar
c$ structure of $J/\psi$ has been recognized almost from very
beginning of its discovery. If it is not a pure $c\bar c$, all the
previous works in terms of the potential models where many
parameters are fixed by fitting data should be re-considered. Or
there may be some other mechanisms which were not taken into
account, or may exist contributions from new physics beyond the
standard model (SM). However, the later seems not very promising
because the concerned energy range is rather low and SM works
perfectly well to explain the data for most states and processes.
Thus it is obviously inclined to the first proposal that $J/\psi$
is not a pure S-wave bound state of $c\bar c$.

It is noted that the calculated results unless for the distribution
$\phi_3$, are one order smaller than the data. The same situation
happens to the $\psi'$, but is still hard to draw a conclusion that
the 14\% rule is due to existence of higher Fock states in $J/\psi$
or other mechanisms which further suppress the reaction of
$\psi'\rightarrow \rho\pi$, may be both. It forms an intriguing
challenge to our understanding of the hadron structures. This whole
picture also applies to $\Upsilon$ family, therefore the future
experiments would provide hints to finally solve the puzzle.

There have been some theoretical explanations besides that we
discussed above, in the work by Mo, Yuan and Wang \cite{X-Mo}, the
authors described the recent status of theoretical research as
well as the experimental measurements on the interesting subject.

Interesting, there is also an alternative opinion towards the
subject, Suzuki \cite{Suzuki}, Zhao \cite{Zhao-Qiang} deny it as a
"puzzle", because they considers that the electromagnetic
interaction may play an important role in $J/\psi$ decays where
$c\bar c$ annihilate into a virtual photon which later fragment
into hadrons. In the picture, it is supposed that a destructive
interference between the contribution of three-gluon and
single-photon processes would suppress $\psi'\to\rho\pi$. Since
its amplitude can be roughly estimated in terms of the measured
rate of $J/\psi$ leptonic decay, there should be a strong
constraint on the proposal. Moreover, if it is true, the
interference would also appear to other decay modes and the
picture should be further investigated and tested by more accurate
experimental data available in the future, especially from the
BESIII.

\section{QCD multi-expansion and strong radiations}\label{sec7}

In an enlightening paper by Kuang \cite{Kuang}, in details, the
author explained the work of Yan  and Kuang \cite{Yan} where they
successfully initiated and developed a complete theoretical
framework, the multi-expansion method in QCD. The theory properly
deals with emission of light hadrons during heavy quarkonia
transitions. Concretely, one mainly studies the processes such as
$\Upsilon(nS)\rightarrow \Upsilon(mS)+\pi+\pi$ or
$\psi(nS)\rightarrow \psi(mS)+\pi+\pi$ with $n>m$. In Ref.
\cite{Kuang}, the author investigated the emission of $h_c$ which
was  found by CLEOc \cite{CLEOc} and intrigued great interests of
theorists \cite{Li-hc} as well as experimentalists \cite{Fang}.
The decay width of such transitions can be written as
$$
\Gamma(n_{I}{}^{3}S_{1}\rightarrow
n_{F}{}^{3}S_{1})=|C_1|^2G|f^{l,P_{I},P_{F}}_{n_{I},l_{I},n_{F},l_{F}}|^2,
$$ where $|C_1|^2$ is a constant to be determined and it comes from
the hadronization of gluons into pions, $G$ is the phase space
factor, $f^{l,P_{I},P_{F}}_{n_{I},l_{I},n_{F},l_{F}}$ is an
overlapping integral over the concerned hadronic wave functions,
their concrete forms were given in \cite{Yan} as
$$
f^{l,P_{I},P_{F}}_{n_{I},l_{I},n_{F},l_{F}}=\sum_K\frac{\int
R_F(r)r^{P_F}R^*_{Kl}(r)r^2dr \int
R^*_{Kl}(r')r'^{P_I}R_I(r')r'^2dr'}{M_I-E_{Kl}},
$$
where $n_{I},n_{F}$ are the principal quantum numbers of initial and
final states, $l_{I},l_{F}$ are the angular momenta of the initial
and final states, $l$ is the angular momentum of the color-octet
$q\bar{q}$ in the intermediate state, $P_{I},P_{F}$ are the indices
related to the multipole radiation, for the E1 radiation
$P_{I},P_{F}$=1 and $l=1$. $R_I,R_F$ and $R_{Kl}$ are the radial
wave functions of the initial and final states, $M_I$ is the mass of
initial quarkonium and $E_{Kl}$ is the energy eigenvalue of the
intermediate hybrid state.

The framework provides a more elegant way to deal with the
long-distance QCD effects even though it only concerns transitions
between states of heavy quarkonia. Moreover, between two gluon
emissions the intermediate state is a hybrid state which
definitely is a subject to draw interests of all high energy
physicists. Just as aforementioned, the QCD theory predicts
existence of exotic states, such as glueball, hybrid, tetraquark
and pentaquark etc., at least does not exclude their existence,
however, so far none of such exotic states have been
experimentally identified yet, so that any direct or indirect
information about the exotic states must be valuable. Since the
subject is heavy quarkonia, the potential model is reasonable to
describe their structures. Recently, many works are devoted to
study the potential which can describe the hybrids which are
composed of heavy quark, anti-quark and a gluon. In this case the
quark and heavy quark reside in a color octet to keep the hybrid
meson in a color singlet, so that the Coulomb potential between
them is repulsive. The recent literature suggests a form by
Swanson and Szczepaniak \cite{Swanson}
$$
V(r)=br+\frac{\pi}{r}\left(1-e^{-fb^{1/2}r}\right),
$$
where the concerned parameters were given in \cite{Swanson}.
Alternatively, Allen et al. proposed another potential form which
includes a repulsive Coulomb piece as \cite{Allen}
$$
V(r)=\frac{\kappa}{8}+\sqrt{(br)^2+2\pi b}+V_0.$$ where $V_0$ is the
zero-point energy and other parameters were also depicted in
\cite{Allen}.

When the works of Refs. \cite{Kuang,Yan} were done, there were not
many data about the transitions available, so that the authors
assumed that $\psi(4.03)$ is the ground state of hybrid $|c\bar
cg\rangle$ and accordingly obtained the concerned parameters in the
potentials listed above. Recently, thanks to the great work of
Belle, Barbar and CLEOc, much more data have been collected and they
enable us to take an inverse strategy to study the problem.

In our strategy, we treat the parameters in the hybrid potential
as free parameters to be determined. Using all the available data
on $\Upsilon(nS)\rightarrow \Upsilon(mS)+\pi+\pi$ and
$\psi(nS)\rightarrow \psi(mS)+\pi+\pi$ as inputs, we apply the
$\chi^2$ analysis with the form of $\bar{\chi}^2$ defined in
\cite{chi} as
$$
\bar{\chi}^2=\sum_i\frac{(W^{th}_i-W^{exp}_i)^2}{(\Delta
W^{exp}_i)^2},
$$
where $i$ represents the $i$-th channel, $W^{th}_i$ is the
theoretical prediction on the width of channel $i$, $W^{exp}_i$ is
the corresponding experimentally measured value, $\Delta
W^{exp}_i$ is the experimental error.

Carrying out all the procedures, we have obtained that the masses
of ground states of the charmonium family and $\Upsilon$ family as
4.23 GeV (for charmonium) and 10.79 GeV (for bottonium). It is
noted that they are not the physical states which are
experimentally observed. It is quite understandable because the
present knowledge may suggest that the gleballs and hybrids may
not exist as real resonances with fixed masses and widths, but mix
with hadrons with regular valence-quark compositions \cite{Close}
and the physical states are the eigenstates of the mass matrices.

However, from other aspects, these results are not accurate due to
large uncertainties of the experimental data. Especially, when we
reached the results, $\Upsilon(5S)$ has not been measured yet, and
we did not include its transition to lower $\Upsilon$ members.

Last year, the Belle Collaboration reported their measurements on
$\Upsilon(5S)\to \Upsilon(1S)\pi^+\pi^-$ and $\Upsilon(5S)\to
\Upsilon(2S)\pi^+\pi^-$ with decays widths
$0.59\pm0.04(\mathrm{stat}.)\pm0.09(\mathrm{syst}.)$ MeV and
$0.85\pm 0.07(\mathrm{stat}.)\pm0.16(\mathrm{syst}.)$ MeV. These
values are about two orders larger than the previously measured
partial widths for dipion transitions between lower $\Upsilon$
resonances \cite{Belle Collaboration}.

Meng and Zhao suggested that the anomalous enhancement is due to
the final state interaction \cite{Meng}. Namely, because
$\Upsilon(5S)$ and $\Upsilon(4S)$ are heavy and above the
production threshold of $B\bar B$, therefore they may first decay
into a $B\bar B$ pair and then by a re-scattering, $B-\bar B$
would turn into $\Upsilon(mS)+\pi\pi$ with $m\leq 3$. By this
picture and fixing the concerned parameters within reasonable
ranges, the enhancement may be understood. If it is true, one
cannot further use the data of $\Upsilon(5S)\rightarrow
\Upsilon(mS)+\pi\pi$ in our above calculations because the
re-scattering contribution contaminates the whole picture and one
is no longer enable to gain direct information about the hybrid
intermediate state at all.

In the calculation, we have found a strange phenomenon that for
$\Upsilon(nS)\rightarrow \Upsilon(mS)+\pi+\pi$, the results are
pretty stable, however for $\psi(nS)\rightarrow \psi(mS)+\pi+\pi$
there exists a cancelation among large numbers with smaller
numbers remaining. This is due to the closeness of the chrmonia
masses and the hybrids, therefore the results on the charmonia
transitions are not very reliable.

On other aspect, such unstableness may be resulted in by a
mistreatment of the charmonia transition. If the final state
interaction is important as the authors of \cite{Meng} suggested,
for bottonia transitions, it would also apply to the charmonia
transition when masses of $\psi(nS)$ are above the production
threshold of $D\bar D$, taking into account such FSI effects, one
may re-extract information about the direct transitions. As did
for the potential model, one may expect that the parameters are
universal for b and c cases and then we can reduce theoretical
uncertainties in the calculation.

Indeed, such information is very necessary for determining the model
parameters and even judge the whole scenario of hybrids. Therefore
we are expecting more data in the charmonia energy regions to be
collected in Babar, Belle and even the LHCb as well as improvements
of theory.

\section{The $X$, $Y$ and $Z$ resonances}\label{sec8}

The QCD theory and quark model have been proven to be very
successful, however, it is by no means the end of the story. Both
QCD and quark model have soft belly where many problems are not
answered yet. Interesting, the two aspects in the quark model and
QCD are connected to each other. In the QCD theory, thanks to the
asymptotic freedom, the perturbation can be trustfully applied to
evaluate any high energy processes, however, on other side, the
low energy processes are governed by the non-perturbative QCD
effects for which so far there lacks any reliable way to deal
with. We have already briefly discussed this issue in previous
section. Unfortunately, many real physics quantities are related
to the low processes, such as the fragmentation in high energy
collisions and hadronic transition form factors in hadron decays.
On other aspect, the quark model demands that mesons are composed
of a pair of quark and anti-quark, baryons consist of three quarks
(or anti-quarks) and QCD interaction (i.e. strong interaction)
binds all the constituents into hadrons. Both of the theories do
not prohibit existence of exotic states, such as glueball, hybrids
and multi-quark states (tetraquark, pentaquark etc.) or even favor
their existence. The pentaquark was a hot topic for a while as
several groups claimed that pentaquark containing an anti-strange
quark or anti-charm quark were observed, then new data of most of
the major labs gave completely negative results. Do they really
exist or mix with regular hadrons as suggested in literature
\cite{Zou}? That is still an unsolved question.

Recently the Babar, belle, CLEO and BES reported many newly
observed resonances which are randomly named as $X$, $Y$ and $Z$
particles. In a review paper, Godfrey and Olsen discussed this
issue in some details \cite{Olsen}. There are many theoretical
works devoting to the exciting field.

\subsection{$D_{sJ}^*(2317)$, $D_{sJ}(2460)$, $D_{sJ}(2860)$ and $D_{sJ}(2715)$}

The discoveries of  mesons $D^*_{sJ}(2317)$ and $D_{sJ}(2460)$
\cite{2317,Belle,CLEO,others} whose spin-parity structure are
respectively $0^+$ and $1^+$, have attracted great interests of
both theorists and experimentalists of high energy physics,
because they seem to be exotic. Bardeen et al. supposed that
$D_{sJ}^*(2317)$ and $D_{sJ}(2460)$ are  $(0^+,1^+)$ chiral
partners of $D_s$ and $D_s^*$ \cite{Bardeen} i.e. p-wave excited
states of $D_s$ and $D_s^*$ \cite{Chaokt}. By studying the mass
spectra,  Beveren and Rupp suggested that $D_{sJ}^*(2317)$ and
$D_{sJ}(2460)$ are made  of $c$ and $\bar{s}$ \cite{Beverenn1}.
With the QCD spectral sum rules, Narison calculated the masses of
$D_{sJ}^*(2317)$ and $D_{sJ}(2460)$ by assuming them as
quark-antiquark bound states and obtained results  consistent with
the experiment data within a wide error range \cite{Narison}. Very
recently, considering the contribution of $DK$ continuum in the
QCD sum rules, Dai et al. obtained the mass of $D_{sJ}^{*}(2317)$
which is consistent with experiments \cite{Daiyb}. Meanwhile, some
authors suggested that $D_{sJ}^*(2317)$ and $D_{sJ}(2460)$ may be
of four-quark structure \cite{Chen,Cheng,T. Barnes}.

Thus to clarify the mist of the structures of $D_{sJ}^*(2317)$ and
$D_{sJ}(2460)$, serious theoretical works are needed. The studies of
the productions and decays of $D_{sJ}^*(2317)$ and $D_{sJ}(2460)$
are very interesting topics. Several groups have calculated the
strong and radiative decay rates of $D_{sJ}^{*}(2317)$ and
$D_{sJ}(2460)$ in different theoretical approaches:  the Light Cone
QCD Sum Rules, Constituent Quark model, Vector Meson Dominant (VMD)
ansatz, constituent quark meson model, etc.
\cite{Nielsen,zhu,decay-1,decay-2,decay-3,decay-4,radi-LCQSR,Liu-2317,wang-2317}.
The authors of Ref. \cite{Cheng,decay-6} also calculated the rates
based on the assumption that $D_{sJ}^{*}(2317)$ and $D_{sJ}(2460)$
are in non-$c\bar s$ structures. Their predictions on the
$D_{sJ}^{*}(2317)$ and $D_{sJ}(2460)$ decay rates are obviously a
few orders larger than that obtained by assuming the two-quark
structure. Recent Faessler et al. study the same subject assuming
$D_{sJ}^*(2317)$ as a $DK$ molecule state using an effective
Lagrangian approach  \cite{Faessler-2317}.

The semileptonic decay of $B_{s}$ is one of ideal platforms to study
the productions of $D_{sJ}^*(2317)$ and $D_{sJ}(2460)$. Especially
the Large Hadron Collider (LHC) will be running in 2008, which can
produce large amounts of  $B_s$. Thus the measurements on $B_{s}\to
D_{sJ}(2317,2460)l\bar{\nu}$ would be realistic. In Ref.
\cite{huang}, author calculated rate of $B_{s}\to
D_{sJ}(2317,2460)l\bar{\nu}$ in terms of the QCD sum rules and HQET.
Recently, authors of Ref. \cite{aliev1,aliev2} completed the
calculations of $B_{s}\to D_{sJ}(2317,2460)l\bar{\nu}$ semileptonic
decays in the QCD sum rules and obtained large branching ratios.
However, the results obtained by authors of Ref.
\cite{aliev1,aliev2} are one order smaller than those given in Ref.
\cite{huang}. In Ref. \cite{semi-2317-liu}, authors studied the same
topic in terms of the Constituent Quark Meson (CQM) model. The
branching ratios of $B_{s}\to D_{sJ}(2317,2460)l\bar{\nu}$ estimated
by the authors of Ref. \cite{semi-2317-liu} and that obtained in
terms of the QCD sum rules \cite{aliev1,aliev2} are of the same
order of magnitude.

In Ref. \cite{liu-psi-2317}, authors use the heavy quark effective
theory (HQET) and a non-relativistic model to evaluate the
production rate of $D_{sJ}^{*}(2317)$ in the decay of $\psi(4415)$,
and find that it is sizable and may be observable at BES III and
CLEO, if it is a p-wave excited state of $D_s(1968)$.

Because $D_{sJ}(2632)$ was only observed by the SELEX
collaboration \cite{selex}, but not by Babar \cite{0408087}, Belle
\cite{2632-belle} and FOCUS \cite{Focus}, its existence is still
in dispute, so we do not intend to include $D_{sJ}(2632)$ in this
review.

In the summer of 2006, the Babar collaboration observed a new
$c\bar s$ state $D_{sJ}(2860)$ with a mass $2856.6\pm1.5\pm 5.0$
MeV and width $\Gamma=(48\pm 7\pm10)$ MeV. Babar observed it only
in the $D^{0}K^{+},D^{+}K_{S}^{0}$ channel and found no evidence
of $D^{\ast 0}K^{+}$ and $D^{*+}K_{S}^{0}$. Thus its quantum
number should correspondingly is $J^{P}=0^{+}, 1^{-}, 2^{+},
3^{-}, \cdots$ \cite{2860-babar}. At the same time, the Belle
collaboration reported a broader $c\bar{s}$ state $D_{sJ}(2715)$
with $J^{P}=1^{-}$ in $B^{+}\to \bar{D}^{0}D^{0}K^{+}$ decay
\cite{belle-2715,Belle-2715}. Its mass is $2715\pm 11^{+11}_{-14}$
MeV and width $\Gamma=(115\pm 20^{+36}_{-32})$ MeV.

According to the heavy quark effective field theory, heavy mesons
form doublets. For example, we have one s-wave $c\bar s$ doublet
$(0^-, 1^-)=(D_{s}(1965),D_{s}^{*}(2115))$ and two p-wave doublets
$(0^+, 1^+)=(D_{sJ}^{*}(2317),\ D_{sJ}(2460))$ and $(1^+,
2^+)=(D_{s1}(2536),D_{s2}(2573))$ \cite{databook}. The two d-wave
$c\bar s$ doublets $(1^-, 2^-)$ and $(2^-, 3^-)$ have not been
observed yet. The possible quantum numbers of $D_{sJ}(2860)$ may
be $0^{+}(2^{3}P_{0})$, $1^{-} (1^{3}D_{1})$, $1^- (2 ^3S_1)$,
$2^+ (2 ^3P_2)$, $2^+ (1 ^3F_2)$ and $3^{-}(1^{3}D_{3})$. The $2
^3P_2$ $c\bar s$ state is expected to lie around $(2.95\sim 3.0)$
GeV while the mass of the $1 ^3F_2$ state will be much higher than
2.86 GeV.

$D_{sJ}(2860)$ was proposed as the first radial excitation of
$D_{sJ}^{*}(2317)$ \cite{beveren-2860,Ma-2860}, or as a
$J^{P}=3^{-}$ $c\bar{s}$ state \cite{colangelo-2860} or  as
$c\bar{s}(2P)$ state \cite{close-2860}. By the potential model, one
can see that $D_{sJ}(2715)$ sits exactly at the position predicted
by the quark model, as 2715 MeV if it is a $2^{3}S_{1}$ $c\bar s$
state \cite{potential model}. The $1^-$ state should lie around
$2721$ MeV if a $(1^+, 1^-$) $c\bar s$  chiral doublet is formed
\cite{chiral}.

In Ref. \cite{Liu-2860}, authors investigated the strong decays of
the excited $c\bar{s}$ states using the $^{3}P_{0}$ model. After
comparing the theoretical decay widths and decay patterns with the
available experimental data, they tend to conclude: (1)
$D_{sJ}(2715)$ is probably the $1^{-}(1^{3}D_{1})$ $c\bar{s}$ state
although the $1^{-}(2^{3}S_{1})$ assignment is not completely
excluded; (2) $D_{sJ}(2860)$ seems unlikely to be the
$1^{-}(2^{3}S_{1})$ and $1^{-}(1^{3}D_{1})$ candidate; (3)
$D_{sJ}(2860)$ as either a $0^{+}(2^{3}P_{0})$ or
$3^{-}(1^{3}D_{3})$ $c\bar{s}$ state is consistent with the
experimental data; (4) experimental search of $D_{sJ}(2860)$ in the
channels $D_s\eta$, $DK^{*}$, $D^{*}K$ and $D_{s}^{*}\eta$ will be
crucial to distinguish the above two possibilities.

In Ref. \cite{Liu-WW-2860}, the strong decay of $D$ wave
$c\bar{s}$ meson to light pseudoscalar meson are studied in the
framework of light cone QCD sum rule (LCQSR).

Recently Dubynskiy and Voloshin proposed an interesting picture to
explain the newly observed rich family of $X$, $Y$ and $Z$. They
suggested that the charmonium states, such as $J/\psi$, $\psi(2S)$,
$\eta_c$, can be bound inside light hadronic matter, especially
inside higher resonances made from light quarks and gluons and they
named such states as hadro-charmonium\cite{Dubynskiy}. Definitely,
this picture should undergo some more serious theoretical and
experimental tests.

Indeed, this is a wide world to be explored, which may help to
testify the quark model and the low-energy QCD principles as well as
all the working phenomenological models.

\section{$D^0-\bar D^0$ mixing}\label{sec9}

This is an extremely interesting subject since a sizable mixing of
$D^0-\bar D^0$ generally implies existence of new physics beyond the
standard model.

In the SM, mixing of particle and anti-particle, such as $K^0-\bar
K^0$, $D^0\bar D^0$ and $B^0_{(s)}-\bar B_{(s)}^0$, occurs via the
box-diagrams \cite{Bigi-BOOK}. The calculation is standard based
on the the GIM mechanism \cite{GIM}. The contribution of the box
diagram is proportional to $m_i^2/m_W^2$ where $m_i$ is the mass
of the exchanged quarks in the box and the CKM matrix elements
\cite{CKM}. Thus for the $B^0-\bar B^0$ and $B_s-\bar B_s$ mixing,
the exchanged quark is the top quark and the factor
$m_t^2/m_W^2>1$ becomes an enhancement, so that the mixing is
large and such mixing has been reported to be observed long time
ago and  in the history it played an important role to hint that
top quark might be heavier than the $W$-boson. However, for
$D^0-\bar D^0$, the exchanged quark can only be b-quark which is
much lighter than the top quark, so that the resultant mixing must
be very small. The $D^0-\bar D^0$ mixing has indeed been measured
by the Babar \cite{Babar-mixing} and Belle \cite{Belle-mixing}
collaborations. Therefore it implies possible existence of new
physics beyond SM. There have been some theoretical suggestion
about the mechanisms which may cause a sizable $D^0-\bar D^0$
mixing, for example, via a FCNC process in the non-universal $Z'$
model \cite{He-mixing} or the unparticle model \cite{Li-mixing}.

The eigenstates of the mass matrix are \cite{Du,Xing}
$$|D_H\rangle=p|D^0\rangle+q|\bar D^0\rangle,$$
$$|D_L\rangle=p|D^0\rangle-q|\bar D^0\rangle.$$
with $|p|^2+|q|^2=1$ and the corresponding eigenvalues are
$$(m_H - m_L)-i (\Gamma_H-\Gamma_L)/2 =2
\sqrt{(M_{12}-i\Gamma_{12}/2)(M_{12}^* - i\Gamma^*_{12}/2)}.$$ where
$M_{12}$ and $\Gamma_{12}$ are the off-diagonal matrix elements and
obtained by calculating the box diagrams, it is noted that both of
them are complex. In SM, they are small, thus a sizable non-zero
mass and life difference of the two physical states $|D_S\rangle$
and $|D_L\rangle$ must be caused by new physics, as aforementioned.

Since the mass and life differences of the two eigenstates are not
very large, it is hard to measure it as one did for the $K^0-\bar
K^0$ system, we can investigate the evolution process of the
$D^0-\bar D^0$ system which in general is produced in $B$-decays
or higher excited states of $\psi$ family. One has \cite{Du}
$$|D^0_p(t)\rangle=g_+(t)|D^0\rangle+{q\over p}g_-(t)|\bar D^0\rangle,$$
$$|\bar D^0_p(t)\rangle={p\over q}g_-(t)|D^0\rangle+g_+(t)|\bar D^0\rangle,$$
and $$g_{\pm}={1\over 2}e^{-imt-{\gamma\over 2}t}\Big[e^{i{\Delta
m\over 2}t-{\Delta\gamma\over 4}t}\pm e^{-i{\Delta m\over
2}t+{\Delta\gamma\over 4}t}\Big],$$ and $\Delta m=m_H-m_L,\;
\Delta\gamma=\gamma_H-\gamma_L$. The important parameters are
$$x={\Delta m\over \gamma},\quad\quad\quad\quad y={\Delta\gamma\over \gamma}.$$ where
$\gamma$ is the average lifetime of $D_H$ and $D_L$.

In analog to the $K$-system, there exists the indirect CP
violation which can be observed in the time evolution of the
system. The direct CP violation will be discussed later.

The data of the Babar collaboration about the $D^0-\bar D^0$
mixing \cite{Babar-mixing} are $x^{\prime2}=[-0.22\pm
0.30(stat)\pm 0.21(syst)]\times 10^{-3}$ and $y'=[9.7\pm
4.4(\mathrm{stat}.)\pm 3.1(\mathrm{syst}.)]\times 10^{-3}$ where
$x'=x\cos\delta_{K\pi}+y\sin\delta_{K\pi}$;
$y'=-x\sin\delta_{K\pi}+y\cos\delta_{K\pi}$ and $\delta_{K\pi}$ is
the strong phase between the Cabibbo favored (CF) and doubly
Cabibbo suppressed (DCS) amplitudes. $x' $is consistent with zero
and $y'$ obviously deviates from zero. The data given by the Belle
collaboration \cite{Belle-mixing} are $y_{CP}=[1.31\pm
0.32(\mathrm{stat}.)\pm 0.25(\mathrm{syst}.)]\%$ where $y_{CP}$ is
defined as $y_{CP}=y\cos\phi-{1\over 2}A_M\sin\phi$. When CP is
conserved, $A_M=\phi=0$, the results are consistent with that
obtained by the Babar collaboration.

The unparticle model was first proposed by Georgi
\cite{Georgi:2007ek}. Georgi argued that operators $O_{BZ}$ made
of BZ fields in the scale invariant sector may interact with
operators $O_{SM}$ of dimension $d_{SM}$ made of Standard Model
(SM) fields at some high energy scale by exchange particles of
large masses, $M_{\cal{U}}$, with the generic form $O_{SM}
O_{BZ}/M^k_{\cal{U}}$. At another scale $\Lambda_{\cal{U}}$ the BZ
sector induces dimensional transmutation, below that scale the BZ
operator $O_{BZ}$ matches onto unparticle operator $O_{\cal{U}}$
with dimension $d_U$ and the unparticle interaction with SM
particles at low energy has the form
$$\lambda \Lambda_{\cal{U}}^{4-d_{SM} -
d_{U}} O_{SM} O_{\cal{U}}.$$ In the SM, the weak phase comes from
the CP phase in the CKM matrix, since the SM contribution to
$M_{12}$ and $\Gamma_{12}$ can be neglected, the weak phase must be
induced in the new physics. We would like to point out some salient
features of the unparticle contribution to $M_{12}^U$ and
$\Gamma_{12}^U$ due to an extra phase factor $e^{-i\pi d_U}$ in the
new scenario. We note that $M_{12}^U$ can have both positive and
negative signs depending on the value of $d_U$ due to the factor
$\cot(\pi d_U)$, therefore if information about the sign can be
obtained from other theoretical considerations or experimental data,
the dimension $d_U$ would be restricted. There may be a sizeable
contribution to $\Gamma_{12}$ at tree level which is not possible
for usual mode where heavy particles are exchanged at tree level.
For $d_{U}$ equal to half integers, there is no contribution to
$M_{12}$, but there is to $\Gamma_{12}$. Another salient feature  is
that the unparticle contribution to the ratio
$M_{12}/(\Gamma_{12}/2)$ is related to the unparticle dimension
parameter $d_U$ by
\begin{eqnarray}
{M_{12}^U\over \Gamma_{12}^U/2} = \cot(\pi d_U).
\end{eqnarray}
If the unparticle contribution dominates meson and antimeson
oscillation then the measurements of $M_{12}$ and $\Gamma_{12}$
would provide a possible way to determine the dimension parameter
$d_U$. In this case, we may gain valuable information about the
dimension of unparticle $d_U$ which so far cannot be directly
obtained when mapping the operator $O_{BZ}$ onto $O_{\cal{U}}$ and
remains as a free and adjustable parameter in most of
phenomenological research works.

Definitely, the contributions of new physics would be added to that
of SM. The new physics effects also contribute to $K^0-\bar K^0$ and
$B^0_{(s)}-\bar K^0_{(s)}$ mixing, however, in those cases the SM
contribution obviously dominates over that from new physics and the
effects of new physics would be smeared out as the measurements are
not very accurate. By contrary, the SM contribution to $D^0-\bar
D^0$ is negligible and all contribution comes from the new physics,
thus measurements on it may provide us with an ideal place for
gaining valuable information about new physics which is indeed the
goal of all high-energy physicists.

To make an accurate measurement, one needs a longer flight time
before the $D$ meson decays, thus the main labs to observe $D^0-\bar
D^0$ would be the two $B$-factories, and the LHCb which will be
running soon will offer us another ideal spot to carry out such
measurements. Even though for this observation, the BESIII does not
have advantages for the kinematics because the linear momenta of the
produced $D$ mesons is small the relativistic time dilation does not
apply,  as the builder promised, the database will be very large and
may greatly enhance the statistics, so that it may also be a
possible lab to observe the mixing effects and explore new physics.
Li and Yang \cite{Yangmz1} studied how to properly extract
information about the mixing from data and make a efficient
analysis.

It is natural to ask if one can observe CP violation in
$D$-system. If $D_H$ and $D_L$ are not CP eigenstates, there could
be an indirect CP violation, however, since $D$ decays faster and
there are many channels available, one cannot determine the
indirect CP violation as easy as in the $K$ system by measuring
$\eta_{+-}$ and $\eta_{00}$ at different distances.

The direct CP violation is defined as
$$C_f(t)=\frac{\Gamma ( D_p^0(t)\rightarrow f)-\Gamma(\bar
D_p^0(t)\rightarrow \bar f)}{\Gamma (D_p^0(t)\rightarrow
f)+\Gamma(\bar D_p^0(t)\rightarrow \bar f)}.$$ which is a
time-dependent measurable quantity. Some details are given in Ref.
\cite{Du}. So far, there is no report on the observation of CP
violation at $D$-system yet, even though the direct CP violation has
been measured to be non-zero at $B$-system. One may expect to make
progress along the line in the future.

\section{Charmed baryons}\label{sec10}

\begin{figure}[htb]
\begin{center}
\scalebox{0.6}{\includegraphics{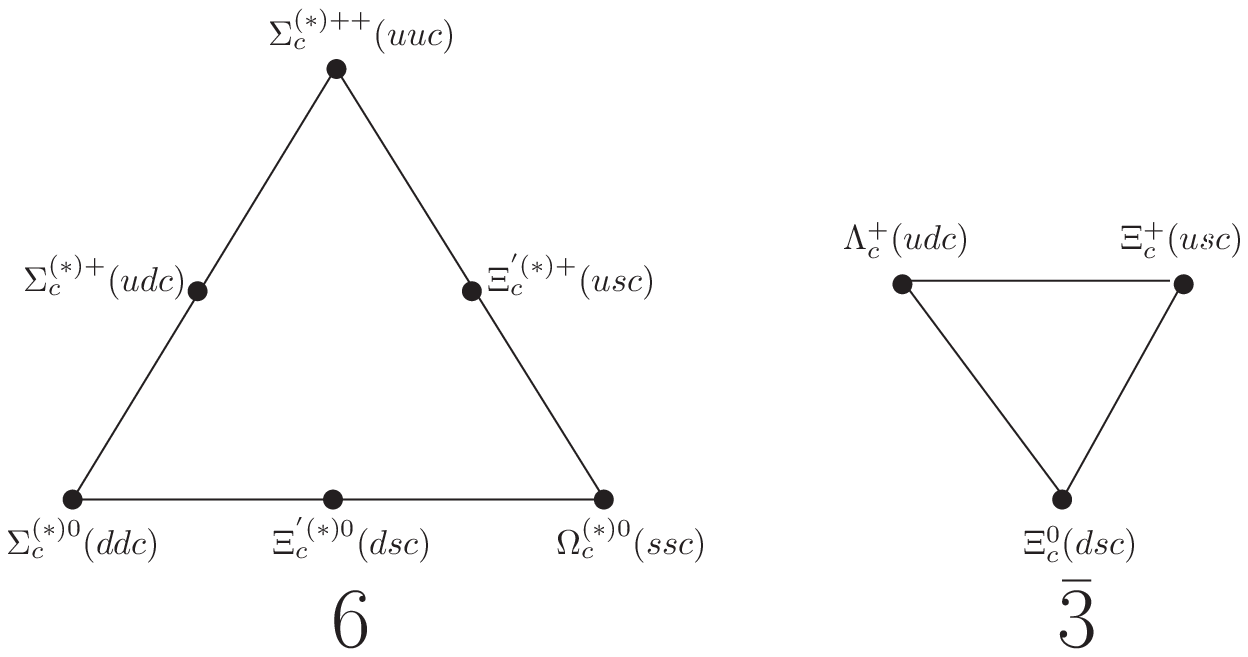}}
\end{center}
\caption{The SU(3) flavor multiplets of charmed
baryons\label{baryon}}
\end{figure}

Let us  present our notations for the excited charmed baryons.
Inside a charmed baryon there are one charm quark and two light
quarks ($u$, $d$ or $s$). It belongs to either the symmetric $6_F$
or antisymmetric $\bar{3}_F$ flavor representation (see Fig.
\ref{baryon}). For the S-wave charmed baryons, the total
color-flavor-spin wave function and color wave function must be
symmetric and antisymmetric respectively. Hence the spin of the two
light quarks is S=1 for $6_F$ or S=0 for $\bar{3}_F$. The angular
momentum and parity of the S-wave charmed baryons are
$J^{P}=\frac{1}{2}^+$ or $\frac{3}{2}^{+}$ for $6_F$ and
$J^{P}=\frac{1}{2}^{+}$ for $\bar{3}_F$. The S-wave charmed baryons
are listed in Fig. \ref{baryon}, where we use the star to denote
$\frac{3}{2}^{+}$ baryons and the prime to denote the
$J^P=\frac{1}{2}^{+}$ baryons in the ${6}_F$ representation.

\begin{figure}[htb]
\begin{center}
\scalebox{0.7}{\includegraphics{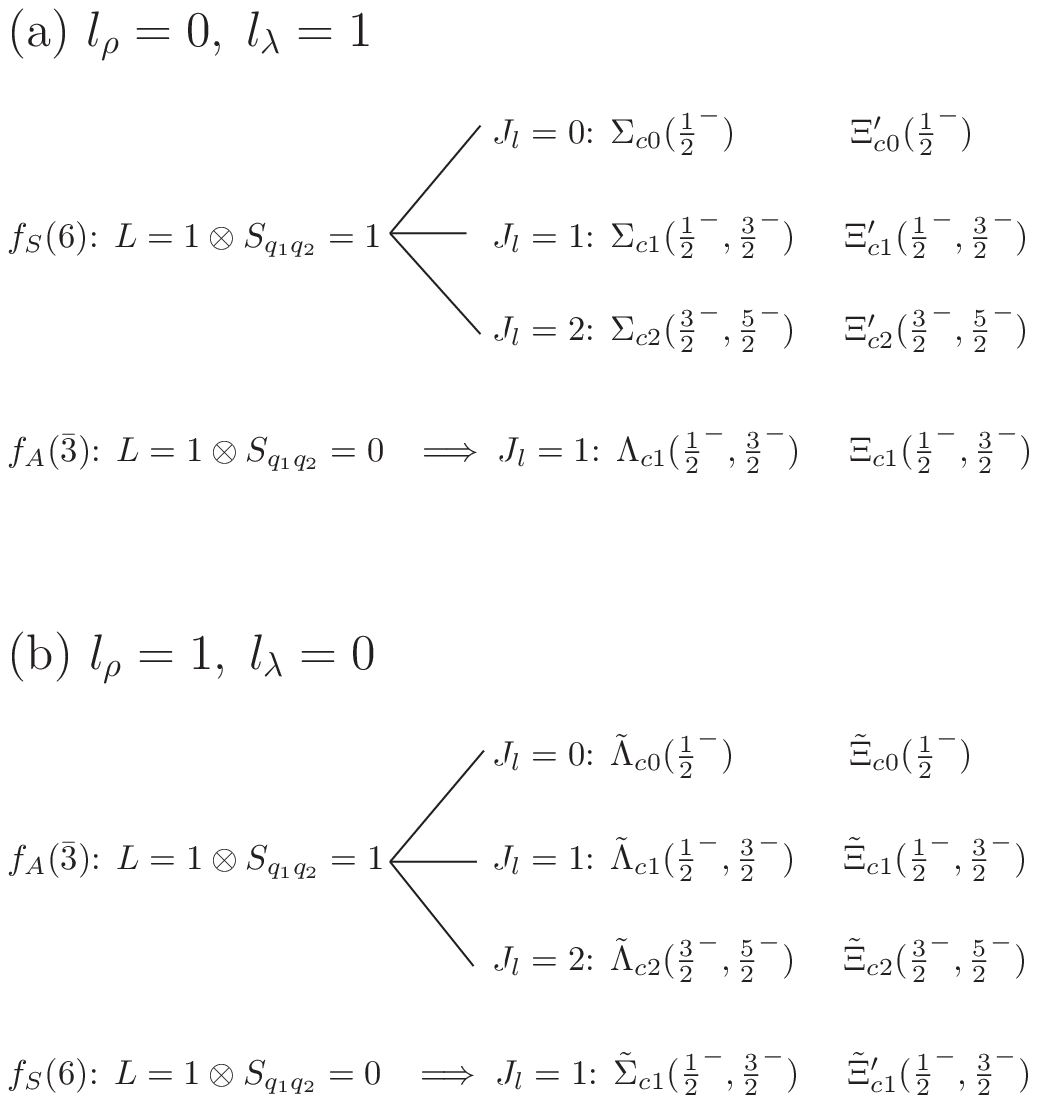}}
\end{center}
\caption{The notations for P-wave charmed baryons. $f_{S}(6_F)$ and
$f_{A}({\bar 3}_F)$ denote the SU(3) flavor representation.
$S_{q_{1}q_{2}}$ is the total spin of the two light quarks. $L$
denotes the total orbital angular momentum of charmed baryon
system.\label{p-wave}}
\end{figure}

In Fig. \ref{p-wave} we present our notations and conventions for
the P-wave charmed baryons. $l_{\rho}$ is the orbital angular
momentum between the two light quarks while $l_{\lambda}$ denotes
the orbital angular momentum between the charm quark and the two
light quark system. We use the prime to label the $\Xi_{cJ_{l}}$
baryons in the $6_F$ representation and the tilde to discriminate
the baryons with $l_{\rho}=1$ from that with $l_{\lambda}=1$.

\begin{center}
\begin{figure}[htb]
\begin{tabular}{ccc}
\scalebox{0.7}{\includegraphics{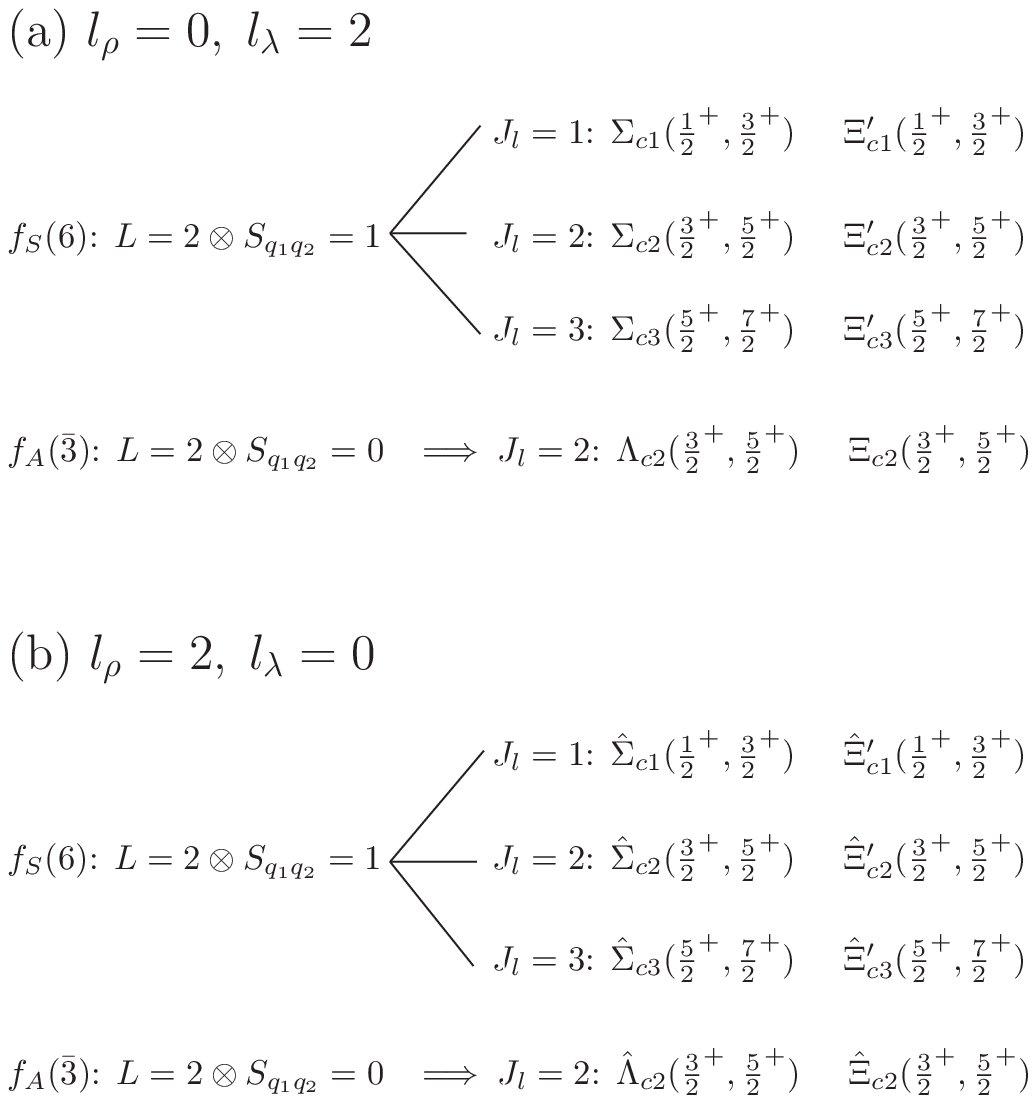}}&
\scalebox{0.7}{\includegraphics{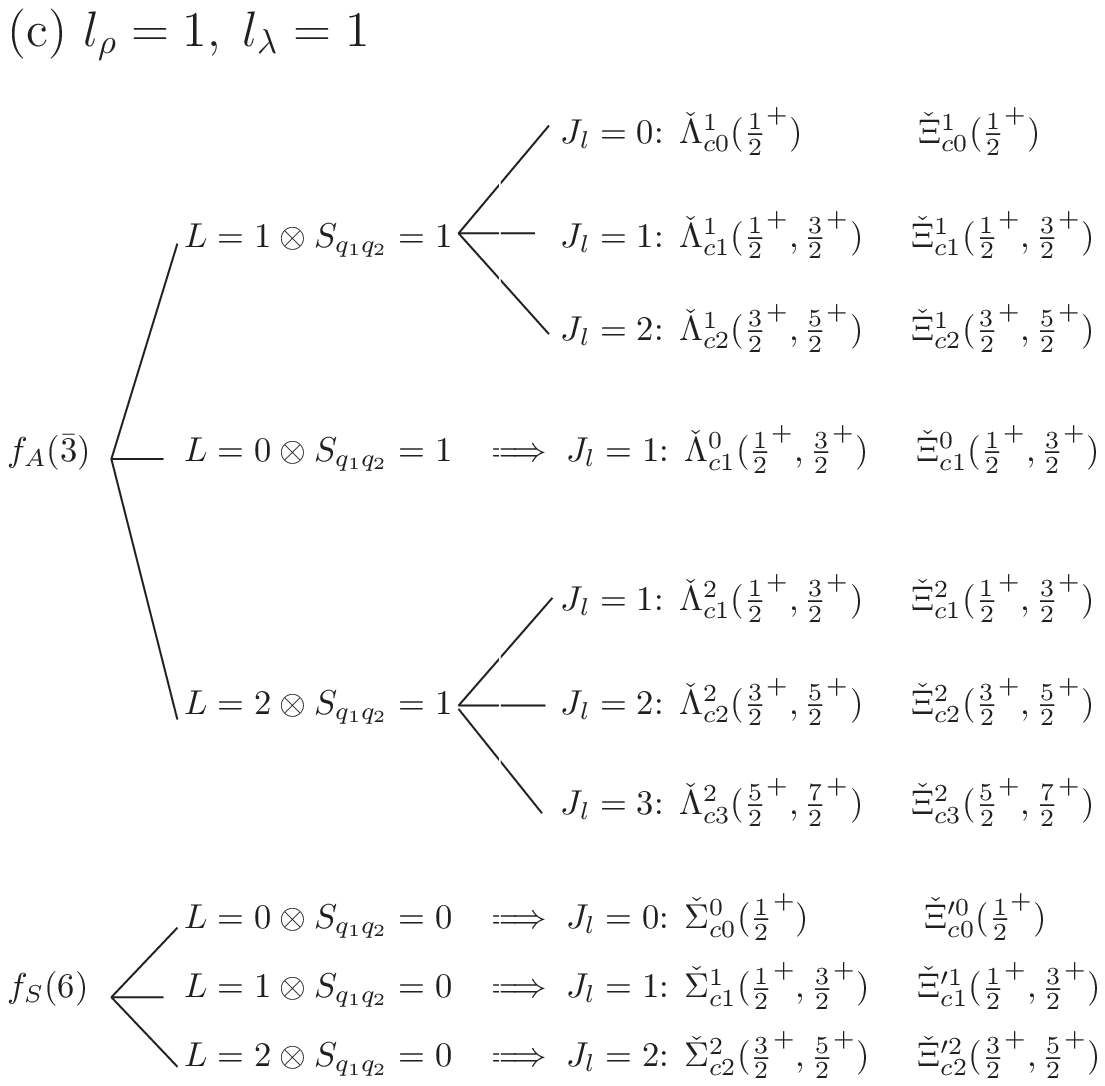}}
\end{tabular}
\caption{The notations for the D-wave charmed
baryons.\label{d-wave}}
\end{figure}\end{center}

The notation for D-wave charmed baryons is more complicated (see
Fig. \ref{d-wave}). Besides the prime, $l_{\rho}$ and $l_{\lambda}$
defined above, we use the hat and check to denote the charmed
baryons with $l_{\rho}=2$ and $l_{\rho}=1$ respectively. For the
baryons with $l_{\rho}=1$ and $l_{\lambda}=1$, we use the
superscript $L$ to denote the different total angular momentum in
$\check{\Lambda}_{cJ_{l}}^{L}$, $\check{\Sigma}_{cJ_{l}}^{L}$ and
$\check{\Xi}_{cJ_{l}}^{L}$.

The Babar and Belle collaborations observed several excited
charmed baryons: $\Lambda_{c}(2880,2940)^+$,
$\Xi_c(2980,3077)^{+,0}$ and $\Omega_{c}(2768)^0$
\cite{babar-2880,belle-2880,babar-2980-3077,
belle-2980-3077,new-Xi3055,babar-omega}.
We collect the experimental information of these recently observed
hadrons in Table \ref{charmed baryon}. Their quantum numbers have
not been determined except $\Lambda_{c}(2880)^+$ \cite{cleo-2880}.
\begin{widetext}
\begin{center}
\begin{table}[htb]
\tiny
\begin{tabular}{c||c|c|ccccccccc} \hline
State&Mass and Width (MeV)&Decay channels in experiments& Other
information\\\hline\hline&$2881.5\pm 0.3$, $<8$
\cite{cleo-2880}&$\Lambda_{c}\pi^{+}\pi^{-}$& \\\cline{2-3}
&$2881.9\pm0.1\pm0.5$ , $5.8\pm1.5\pm1.1$ \cite{babar-2880}&$D^0 p$&
\raisebox{2ex}[0pt]{$J^{P}$ favors $\frac{5}{2}^+$
\cite{belle-2880},}\\\cline{2-3}\raisebox{3ex}[0pt]{$\Lambda_{c}(2880)^{+}$}
&$2881.2\pm0.2^{+0.4}_{-0.3}$,
$5.5^{+0.7}_{-0.5}\pm0.4$ \cite{belle-2880}&$\Sigma_{ c}^{\star
0,++}(2520)\pi^{+,-}$&\raisebox{2ex}[0pt]{$\frac{\Gamma(\Sigma_{
c}^{\star}(2520)\pi^{\pm})}
{\Gamma(\Sigma_{c}(2455)\pi^{\pm})}=0.225\pm0.062\pm0.025$
\cite{belle-2880}}
\\\hline\hline
&$2939.\pm1.3\pm1.0$, $17.5\pm5.2\pm5.9$ \cite{babar-2880}&$D^0 p$&\\
\cline{2-3}\raisebox{2ex}[0pt]{$\Lambda_{c}(2940)^+$}&$2937.9\pm1.0^{+1.8}_{-0.4}$,
$10\pm4\pm5$ \cite{belle-2880}&$\Sigma_{c}(2455)^{0,++}\pi^{+,-}$
&\raisebox{2ex}[0pt]{-}\\\hline\hline &$2967.1\pm1.9\pm1.0$,
$23.6\pm2.8\pm1.3$ \cite{babar-2980-3077}&$\Lambda_{c}^{+}K^{-}
\pi^{+}$&\\\cline{2-3}
\raisebox{2ex}[0pt]{$\Xi_{c}(2980)^+$}&$2978.5\pm2.1\pm2.0$,
$43.5\pm7.5\pm7.0$ \cite{belle-2980-3077}&$\Lambda_{c}^{+}K^{-}
\pi^{+}$&\raisebox{2ex}[0pt]{-}\\\hline\hline
$\Xi_{c}(2980)^0$&$2977.1\pm8.8\pm3.5$, $43.5$
\cite{belle-2980-3077}&$\Lambda_{c}^{+}K_{S}^{0}
\pi^{-}$&-\\\hline\hline &$3076.4\pm0.7\pm0.3$, $6.2\pm1.6\pm0.5$
\cite{babar-2980-3077}&$\Lambda_{c}^{+}K^{-}
\pi^{+}$&\\\cline{2-3}
\raisebox{2ex}[0pt]{$\Xi_{c}(3077)^+$}&$3076.7\pm0.9\pm0.5$,
$6.2\pm1.2\pm0.8$ \cite{belle-2980-3077}&$\Lambda_{c}^{+}K^{-}
\pi^{+}$&\raisebox{2ex}[0pt]{-}\\\hline\hline
$\Xi_{c}(3077)^0$&$3082.8\pm1.8\pm1.5$, $5.2\pm3.1\pm1.8$
\cite{belle-2980-3077}&$\Lambda_{c}^{+}K_{S}^{0}
\pi^{-}$&-\\\hline\hline $\Xi_{c}(3055)^+$
\cite{new-Xi3055}&$3054.2\pm 1.2\pm 0.5$, $17\pm6\pm11$&
$\Lambda_c^+ K^-\pi^+$&-\\\hline\hline $\Xi_{c}(3122)^+$
\cite{new-Xi3055}&$3122.9\pm 1.3\pm 0.3$, $4.4\pm3.4\pm1.7$&
$\Lambda_c^+ K^-\pi^+$&-
\\\hline\hline
 $\Omega_{c}(2768)^{0}$&$2768.3\pm 3.0$
\cite{babar-omega}&$\Omega_{c}^{0}\gamma$&$J^{P}=\frac{3}{2}^{+}$\\\hline
\end{tabular}
\caption{A summary of recently observed charmed baryons by Babar and
Belle collaborations. \label{charmed baryon}}
\end{table}
\end{center}
\end{widetext}

In the past decades, there have been some research works on heavy
baryons \cite{charmed baryons,review-charmed,review-charmed-1}.
However these new observation inspired new investigations of these
states
\cite{charmed-baryons-rosner,charmed-baryons-xiang,charmed-baryons-cheng,
charmed-baryons-valcarce}.
In Ref. \cite{charmed-baryons-xiang}, authors studied the
$\Lambda_{c}(2940)^+$ and its possible decay modes assuming
$\Lambda_{c}(2940)^+$ to be a $D^{*0}p$ molecular state
\cite{charmed-baryons-xiang}. Cheng et al. calculated the strong
decays of newly observed charmed mesons in the framework of heavy
hadron chiral perturbation theory (HHChPT)
\cite{charmed-baryons-cheng}. In order to understand their
structures using the present experimental information, in Ref.
\cite{charmed baryons-liu,charmed baryons-liu-1}, the strong decay
pattern of the excited charmed baryons are studied systematically in
the framework of the $^{3}P_{0}$ strong decay model. After comparing
the theoretical results with the available experimental data, their
favorable quantum numbers and assignments are obtained in the quark
model.

It is an interesting field. There was a puzzle that the lifetime
of $B^0$ is close to that of $B^{\pm}$ ($\tau(B^0)=(1.530\pm
0.009)\times 10^{-12}\ s,\;\;\tau(B^{\pm})=(1.638\pm 0.011)\times
10^{-12}\ s$) whereas the lifetime of $D^{\pm}$ is almost double
of that of $D^0$ ($\tau(D^0)=(410.1\pm 1.5)\times 10^{-15}\
s,\;\;\tau(D^{\pm})=(1040\pm 7)\times 10^{-15}\ s$). By the heavy
quark effective theory, the total width of a meson is mainly
determined by the total width of the heavy quark constituent in
the meson which is proportional to $m_Q^5\times V_{CKM}$ where
$V_{CKM}$ is the corresponding CKM matrix elements. Obviously, the
weak decays of b-quark is Cabibbo suppressed while for charm
quark, it is Cabibbo favored. The obvious difference between $B$
and $D$ lifetimes is excellently explained by Bigi et al.
\cite{Bigi}. Another puzzle is the lifetime difference between
$\Lambda_b$ and $B$ meson, which used to be as large as 0.29,
however the recent measurement alleviates the discrepancy as
$\tau(\Lambda_b)/\tau(B^0) =1.041\pm 0.057$ \cite{Lambdabexp}, and
is close to the theoretical evaluation \cite{Gabbiani}. However,
in the theoretical calculation one needs to evaluate the hadronic
matrix element in analog to the discussion given in previous
sections about the meson case, and the estimate still has a large
uncertainty, so  we do not think that the problem is completely
solved.

More intriguing issue is that the lifetime of $\tau(\Lambda_c)
(\approx (200\pm 6)\times 10^{-15}\ s$) is much smaller than that
of $D^0$ and $D^{\pm}$, and it implies that the light quark cannot
completely be treated as  spectators and the binding effects may
also be important. In the heavy quark language, the ${1\over
m_c^n}\;(n\geq 1)$ corrections may play important roles when the
total width is evaluated. The question is that for exclusive
decays, can one use the pQCD or similar method where the $1/m_Q$
expansion is adopted? It seems that one can apply the pQCD method
to analyze the $\Lambda_b$ decays \cite{Guo-peng}, but so far,
there is not work devoting to pQCD application in calculating
$\Lambda_c$ decays yet. The reason is that the energy scale of
$M_{\Lambda_c}$  is relatively low and the exchanged gluons in the
processes are not hard enough to enable us to consider only the
leading, at most the NLO contributions. How to involve more higher
terms in the expansion may be a very challenging task for
theorists.

A great progress in the field is that the double-charmed baryon
$\Xi_{cc}$ was observed by SELEX \cite{SELEX-DOUBLE
CHARM,SELEX-DOUBLE-1,SELEX-DOUBLE-2,SELEX-confirm-double,SELEX-new
CHANNEL} at the Fermi Lab. Since it contains two charm quarks which
would decay via weak interaction independently, one may expect to
gain some information about the binding effects. In analog with the
method given by Bigi et al. \cite{Bigi}, we evaluated
\cite{lifetime} the lifetimes of $\Xi_{cc}^+,\; \Xi_{cc}^{++}$ and
$\Omega_{cc}^+$ which contain different light flavors, to order of
$1/m_c^3$. The hadronic matrix elements were evaluated in terms of
the simplest model, i.e. the harmonic oscillator model
\cite{harmonic}. The theoretical predictions can be tested in the
future experiments especially the LHCb. Indeed, we lay our hope on
the LHCb data which may produce a large database on such doubly
charmed baryons, and maybe also baryons containing two b-quarks
$\Xi_{bb}$, or b and c quarks $\Xi_{bc}$ etc. By studying them, we
will gain valuable information on the non-perturbative QCD effects
which bind the heavy quarks together with a light flavor. Moreover,
one may expect to observe baryons which are composed of three heavy
quarks ($b$ and/or $c$). There has been some theoretical works in
this field \cite{Brambilla}, but phenomenological studies on their
production rates and decay modes require more serious works in order
to investigate their structures and governing mechanisms in some
details. Some details of  baryons containing two heavy quarks have
been reviewed  by Kiselev and Likholded  in an enlightening paper
\cite{Kiselev}.

We have also proposed a special decay mode of $\Xi_{cc}$ to
investigate new physics. Namely, we suggest to measure a direct
decay of $\Xi_{cc}$ into non-charm final states \cite{double-charm},
which is definitely a signal for new physics beyond SM. Concretely,
in the SM, the direct decay of $\Xi_{cc}$ into non-charm final
states are very restricted and the sequential decays, such as
$\Xi_{cc}\rightarrow \Lambda_c+\bar D\rightarrow \Lambda+K+X$ would
produce final states which  involve at least more than two hadrons,
so that can be easily distinguished from the expected decay modes
with two non-charmed hadrons. We have tried to employ two models
which recently become more popular to the theorists, the $Z'$ and
unparticle models which can induce the flavor-changing neutral
current at tree level, so can realize the reaction $cc\to qq'$ where
$q$ and $q'$ are light flavors. However, our numerical results show
that the two models cannot cause sizable fractions for practical
measurement, even though one can have a great luminosity at LHCb.
Our conclusion is that if such decay modes are observed, one can
claim that as a definite evidence of new physics, but it is neither
the $Z'$ model, nor the unparticle model. This would motivate us to
explore other new physics models beyond the SM.

The charmed pentaquark was a hot topic and the QCD theory does not
exclude  pentaquark, the question is do we need to take it more
seriously and where should we search for them and how to identify
them from the regular baryons \cite{Ma-1}. As we discussed in
previous sections, many theorists favor tetraquark as a plausible
explanation of some newly observed $X$, $Y$ and $Z$ resonances,
therefore why should we firmly reject existence of pentaquark? A
full scan of the pentaquark is not intended in this review work,
but we would like to analyze some aspects concerning the charm
physics. The first proposal on existence of  pentaquark is based
on the group theory analysis \cite{helima}, that the pentquark is
composed of four $u$ and $d$ quarks and one anti-strange quark
with both the strangeness baryon number being +1. Then a claim of
finding charmed pentaquark was made where a $\bar c$ replaced
$\bar s$. However, later most major labs in the world reported
negative results in search for pentaquark and the society of high
energy physics tends to deny discovery of pentaquark. We also made
effort to look for trace of pentaquark via some processes, such as
photo-production and radiative decay of pentaquark as well as the
hadronic decays involving pentaquarks \cite{He-pentaquark}, and
the conclusion is still that more accurate measurements are
necessary. On other aspect, we seriously consider that the
pentaquark may mix with the other baryonic states
\cite{He-mixing-pen}.

It is interesting to investigate the quark-structure of pentaquark
which is favored by QCD. Besides assuming that all the four quarks
and an anti-quark all mix together in a big hadronic bag, the more
favorable structures are proposed by several authors, for example,
Jaffe and Wilczek \cite{Jaffe} suggested the
diquark-diquark-anti-quark structure whereas Karliner and Lipkin
\cite{Lipkin} favored the diqaurk-triquark configuration. The key
point is why the pentaquark has not been observed in any colliders
but was seen in fixed target experiments. Lipkin tried to explain
the situation by considering the initial quark configuration in
the beam particle or target. This should draw serious attentions
of experimentalists and theorists and design new experiments to
search for evidence of pentaquarks. Since in the $p\bar p$
collisions at LHC, the rich quark contents and high luminosity may
be a good source to produce pentaquarks, and until then we can
draw a definite conclusion if the pentaquark can exist as a real
particle. In this line the charmed pentaquark may be more
favorable because it contains a heavy flavor. The most interesting
subject would be that if the pentaquark is strictly prohibited,
there must be a symmetry to restrict it, it would require a new
understanding on the mechanism beside the general theory of QCD.

\section{Diquark}\label{sec11}

It is an extremely interesting topic, not only for charmed
baryons, but since it is widely applied in studying physical
processes where baryons are involved, diquark is worth careful
investigation. By the SU(3) theory, two quarks can reside in a
color-anti-triplet $\bar 3$ to constitute a loose bound state and
then it combines with the rest quark to make the baryon of color
singlet. The QCD induced potential is proportional to the Casimir
factor $\langle\psi|\lambda^a\lambda^a|\psi\rangle$ where
$\lambda^a$ is the SU(3) generator and $|\psi\rangle$ is the
state, thus in a color-anti-triplet (or a triplet for two
anti-quarks) the two quarks attract each other because the Casimir
factor is negative. It seems that the bound state is plausible,
but for light quarks, their linear momenta in the diquark are
rather large and may spread out in space, thus cannot make a real
physical subject in common sense. Therefore whether the diquark
structure is only a mathematical description for a baryon or it
can be a physical subject is still in dispute. That is how to
understand the subject "diquark" is an intriguing topic in
theoretical physics.

However, if the baryon contains a heavy quark, it becomes simpler
as usually assumed, the heavy quark may sit near the geometric
center of the hadron (almost at the center) and one can further
postulate that the linear momentum of the heavy quark is zero.
Then the two light quarks would compose a diquark even though its
spatial spreading is still large. Therefore when we apply the
concept of diquark in this case, we indeed do not treat it as a
point-like particle, but a loosely bound state with a common
momentum and interact with the heavy quark as a whole object,
moreover, when some reaction takes place, it may also perform as a
whole object, but to compensate the effects of its inner motion,
one needs to introduce a form factor(s) whose phenomenological
form was given in Ref.{anselmino1} as
$F(Q^2)=\frac{Q^2_0}{Q^2_0+Q^2}$, where $Q_0$ is a parameter and
is determined as   $Q_0^2\sim 3.2$ GeV$^2$ by fitting data
\cite{anselmino1}. The form is obviously understandable. The form
factors should be normalized to unity as $Q^2\to 0$, i.e. as one
looks at the diquark from a far distance, the form factor becomes
a unity, whereas as $Q^2\to\infty$, the inspector then penetrates
into the diquark, so that he would see the individual quarks
instead of the whole and the diquark picture no longer holds and
mathematically it is required to approach zero as $Q^2\to \infty$.
The form factors should be calculated in terms of a more
fundamental way, i.e. based on quantum field theory. However, the
non-perturbative QCD effects can by no means be treated completely
based on an underlying theory so far, instead we need to invoke
phenomenological models. The Bethe-Salpeter equation obviously is
a reasonable approach where the kernel depends on concrete models.
The form factor obtained in terms of the BSE generally coincides
with the picture described above. Guo, Thomas and Williams
\cite{Guo} studied the $1/m_Q$ corrections for the B-S equations
for $\Lambda_Q$, $\Omega_Q$ in the diquark picture. Then in the
framework of the B-S equation, we calculate the form factors for
various types of diquarks \cite{Ke} and the form factors obtained
based on the B-S theory are qualitatively consistent with that
given in Ref. \cite{anselmino1}. Applying the diquark picture, we
calculate several processes where heavy baryons participate
\cite{Dai}. The results are somehow competent to be compared with
data, and it indicates that the diquark picture is vigorous and
robust, even though still needs further studies.

It may be worth mentioning a special situation, namely the baryons
containing two heavy quarks should fit in the diquark picture well.
Falk et al. \cite{Falk} considered the two charm quarks in a doubly
heavy baryon constitute a "perturbatively bound diquark" whose
wavefunction at origin follows $$|R_{(cc)}(0)|^2\approx {1\over 8}
|R_{\psi}(0)|^2\approx (0.41\ GeV)^3,$$ and the factor 1/8 is due to
the difference of the color structures of color anti-triplet and
singlet. Georgi and Wise proposed a very special symmetry called as
the superflavor symmetry \cite{Georgi-Carone}. Simulating the
supersymmetry where the SM fermions have their supersymmetric scalar
partners and the SM bosons also have their SUSY fermionic partners,
in the superflavor symmetry, the heavy quark is written in a doublet
with its superflavor partner (scalar or vector) of color triplet as
$$\Psi_s=(h_v^+, \chi_v)^T;$$ and $$\Phi_v=(h_v^+, A^{mu}_v)^T,$$
where $h_v^+$ is a heavy quark and obeys $\rlap /vh_v^+=h_v^+$,
$\chi_v$ and $A^{\mu}_v$ are  heavy color triplet scalar and
vector respectively, with constraint $v_{\mu}A^{\mu}_v=0$. With
this symmetry, we can estimate the production rates of heavy
baryons which include two heavy quarks, i.e. a heavy diquark. In
this scenario, the heavy scalar or vector diquarks can be treated
as the superflavor symmetric partners of heavy quark and then one
can compare the baryon processes with corresponding meson cases
where two heavy mesons are produced. For both cases of baryon and
meson, the light quark would play the same role. In the meson
cases, at the leading order, there is only one Isgur-Wise function
to manifest the non-perturbative QCD effects, thus by the
superflavor symmetry, we would also apply the same function
\cite{Guo-Jin-Li} to calculate the production rates of $X_{cc},\
X_{bc}, \ X_{bb}, \ etc.$ in $e^+e^-$ collisions. Definitely the
same procedure can be applied to calculate the heavy baryon
production rates at LHC, even though the situation is a bit more
complicated.

\section{Production of charm}

Production of charm, i.e. production of charmed mesons, baryons and
charmonia, is a large subject, and we are not going to cover this
field in this short review, but indicate some tricky and challenging
issues.

The subject of production of $J/\psi,\ \psi' \:{\rm and }\:
\Upsilon$ was reviewed by Lansberg \cite{Lansberg} in some details.
In this incomplete review, we do not intend to cover the whole
subject, but indicate that this field is still tricky and intriguing
by two examples which are widely discussed in the society.

As a long known question, the inclusive $J/\psi$ production in
$e^+e^-\rightarrow J/\psi c\bar c$ raises a challenge to our theory.
At LO, the QCD prediction on the cross section is smaller than the
observed value by a factor of 5. This discrepancy may come from the
wavefunctions of the produced $J/\psi$ \cite{Ebert} or the NLO
corrections. In an enlightening work \cite{Chao}, the authors
suggested the NLO may enhance the cross section by a factor about
1.8. This may greatly change the situation.

However another serious question is raised that if the NLO can
change this cross section so much, what mechanism or symmetry
determine this unusual enhancement. Generally the loop suppression
is proportional to $\alpha_s/\pi$ and at this energy scale
$\alpha_s$ is smaller than 0.2. If the number of diagrams (at NLO,
the number of diagrams is rather large) is the reason as long as the
interference among the diagrams are mostly constructive, one could
ask how about the contribution of NNLO? Definitely, even though the
diagrams at NLO are constructive, it does not suggest that they are
constructive at NNLO at all. If the NNLO contribution decreases, we
can happily declare that the theory works well, however if it
continues to increase or the correction at NNLO is even larger than
that of NLO, the unitarity would confront a serious challenge and
one needs to take the hint more seriously. Indeed, to give a firm
answer to the question, one must calculate the NNLO corrections.
However, there are over 1000 two-loop diagrams and a complete
calculation is extremely difficult project. Since it is necessary,
some serious work must be done and it is indeed worthwhile.

The problem about the production of $J/\psi$ was investigated by
several groups \cite{Lansberg-1}. The authors studied the
hadronprocution of $J/\psi, \ \psi',\ \Upsilon$ in associated with
heavy quark pair which was the subject we listed above and
investigated further, and a possible solution of the $J/\psi$
production puzzle. Their result was confirmed by  Gong and Wang
\cite{Wangjx}. In the paper \cite{Wangjx}, the authors further
studied the $p_t$ distribution of the $J/\psi$ polarization in
$pp\rightarrow J/\psi+X$ process. There is also serious discrepancy
between theoretical prediction and data. They calculated the NLO
$p_t$ distribution of $J/\psi$ polarization and found an amazing
change. Namely, they find that the polarization status of $J/\psi$
at large $p_t$ changes from the transverse-polarization dominance at
LO into the longitudinal polarization dominance at NLO. This
correction indeed partly makes up the gap between theoretical result
and data. In their first work, they only considered the
color-singlet contribution and got improvement, so there was a hope
that when the color-octet contribution was involved the situation
would be improved further and the data could be explained
eventually. But unfortunately, when the color-octet contribution is
taken into account, the new correction is not so drastic and does
not change the whole picture much, therefore the large gap between
theoretical prediction and data still exists. The authors
\cite{Wangjx} claimed that to explain the data, new mechanisms or
new physics or new understanding of the $J/\psi$ structure etc. are
needed.

The two examples do not intend to cover the large-scale field of
production of charm, but indicate that there are still many problems
which are not fully understood yet, a lot of theoretical efforts are
required. This is also associated with the puzzle we  discussed in
this review that why pentaquark has not been observed and so on.

\section{Discussion}

This is by no means a definite conclusion because in this review,
we raise many unsolved problems and list some references where
many authors devote great efforts in this fascinating field. The
theoretical difficulty is obvious that charm quark is heavy, but
not sufficiently heavy, so non-perturbative QCD effects still play
important roles, and moreover, the relativistic effects of quark
quark in hadrons are also not negligible. Therefore the heavy
quark effective theory (HQET) may apply to this field, but needs
to consider higher orders in the $1/m_c$ expansions and it brings
unexpected uncertainties. The good example is the lifetime
difference between $\Lambda_c$ and $D$ mesons, compared to the
$B$-cases. On other aspect, careful studies in the charm field may
greatly enrich our understanding of the basic principles,
concretely, how to properly deal with the non-perturbative QCD
effects is one of the  most important issues. Moreover, the work
by the authors \cite{Chao,Wangjx} indicates that higher order QCD
corrections may be very important, and even play a crucial role.
This indeed further intrigues the field of charm physics.

The main task in this field seems not closely related to searching
for new physics beyond SM, as generally considered because of its
energy scale, but there indeed is possibility to investigate new
physics. For example, the $D^0-\bar D^0$ mixing as aforementioned,
and the lepton flavor violation processes besides the neutrino
oscillation experiments. Namely, one can examine the reactions like
\cite{Xu-Ye}
$$e^+e^-\rightarrow \psi\rightarrow e^{-(+)}\mu^{+(-)},$$ for which
the charm-tau energy region may be the ideal observation place. On
the theoretical aspect, in the SM, this FCNC can occur at one-loop
order where neutrino masses need to be large enough to produce
observational effects. However, it contradicts to all the data of
the solar neutrino, atmospheric neutrino experiments and one can
claim that observation of such reaction is a clear signal of new
physics beyond SM. There are so many new physics models which can
produce this reaction, can the reaction, if it indeed occurs,
distinguish among them? Generally it cannot. But one can at least
tell which model is a possible one and maybe, such a reaction would
help to get rid of a few proposed models.

There are general discussions on possibilities to measure the
production rates from the theoretical angle \cite{Zhangxm} and set
bounds on the model parameters. Very recently, we suggest
\cite{Wei-Xu-Li} that the unparticle model may induce such FCNC. By
adopting the reasonable region of the concerned parameters, our
calculation shows that one can expect a large production rate near
the resonance peak of $J/\psi$. Carefully analyzing possible
background, one has confidence to extract clear signal for the
luminosity of PEPC II and detection accuracy of BESIII. If it is
observed, the unparticle model at least is one of favorable
candidates of new physics, on other side, if it is not observed, the
unparticle model would be somehow disfavored.

There are too many questions to be answered in this energy region,
as we discussed above, there are several very enlightening review
papers which shed lights on this exciting field
\cite{Yalsley,Voloshin,Olsen1,Valcarce}.

Indeed, besides the major line, there are some other subjects which
should be researched in this field, for example, it is proposed to
test the Bell inequality in high energy processes,
$\eta_c\rightarrow \Lambda\bar\Lambda\rightarrow p\pi^-+\bar p\pi^+$
\cite{Tornqvist} and decays of charmonia into kaon \cite{Bramon}. As
a matter of fact, the scheme proposed in \cite{Tornqvist} was
carried out in terms of the DM2 data, however, because the database
was too small, the poor statistics made the work meaningless. Today,
the situation will be different, the BESIII will provide a
tremendously large database of $J/\psi$, so that one may expect to
make accurate measurements to testify the Bell inequality.

We believe that we are standing at a prosperous epoch for high
energy physics. The two B-factories continue to produce more data
where one may also have a chance to study charm-related physics,
such as CP violation and higher excited states of charmonia and
charmed mesons and baryons. Moreover, the LHC will be running soon
and not only a lot of information towards underlying physics such
as the Higgs mechanism and new physics beyond SM will be
collected, but also some details about bottom and charm physics
can be achieved, especially we may lay great hope on the LHCb.
Even the ALICE can tell us some interesting subjects about charm
physics, for example, if the $J/\psi$ suppression in
quark-gluon-plasma (QGP) really exists, can the higher temperature
and pressure change the potential between quarks, can we expect an
observable phase transition, like deconfinement and chiral
symmetry restoration etc. More than anything else which can enrich
our knowledge on charm physics, the BEPC II and BESIII will be
operating very soon, and a great database on $J/\psi$ and other
members of the $\psi$ family, as well as the D-mesons, baryons,
even tensors will be available, so that we are sure, may new
challenges will be waiting for us. So we feel much encouraged than
any past time, because so many questions need to be
solved and more data will help to do the job.\\

\noindent Acknowledgement: This work is supported by the National
Natural Science Foundation of China and the Special Grant of the
Education Ministry of China. We thank S. Olsen, C.H. Chang, K.T.
Chao, B.S. Zou, S.L. Zhu, Q. Zhao, S. Jin, X. Shen, C. Yuan, T. Li,
H.W. Ke and many others for helpful discussions. We sincerely
apologize that our reference list is very incomplete, so that many
important works may be missed because of our limited reading. One of
authors (X.Liu.) would like to thank  the \emph{Minist\'{e}rio da
Ci\^{e}ncia, Tecnologia e Ensino Superior} \/of Portugal for
financial support under contract SFRH/BPD/34819/.


\baselineskip 22pt


\begin{thebibliography}{99}
\bibitem{unitarity} M. Veltman, {\it Facts and Mysteries in elementary Particle
physics}, World Scientific Publishing Co.Pte. Ltd. (2003).

\bibitem{anomaly} D. Gross and R. Jakiew, Phys. Rev. {\bf D 6},
477 (1972); H. Georgi and S. Glashow, Phys. Rev. {\bf D 6}, 429
(1972).

\bibitem{Gaillard} M.K. Gaillard and B.W. Lee, Phys. Rev. {\bf D
10}, 897 (1974).

\bibitem{Ting-Richter}J. Aubert et al., Phys. Rev. Lett. {\bf 33}, 1404 (1974);
J. Augustin et al., Phys. Rev. Lett. {\bf 33}, 1406 (1974).

\bibitem{TEVATRON}CDF Collaboration, F. Abe et al., Phys. Rev. Lett.
{\bf 74}, 2626 (1995); D0 Collaboration, S. Abachi et al., Phys.
Rev. Lett. {\bf 74}, 2632 (1995).

\bibitem{potential}C. Quigg and J.L. Rosner, Phys. Rep. {\bf 56},
167 (1979); Phys. Rev. {\bf D 23}, 2625 (1981); C. Quigg, H.B.
Thacker and J.L. Rosner, Phys. Rev. {\bf D 21}, 234 (1980).

\bibitem{Voloshin}M.B. Voloshin, arXiv:0711.4556 [hep-ph].

\bibitem{rhopi-exp}Mark II Collaboration, M. Franklin et al.,
Phys. Rev. Lett. {\bf 51}, 963 (1983); BES Collaboration, J. Bai
et al., Phys. Rev. {\bf D 69}, 072001 (2004); CLEO Collaboration,
N. Adam et al., Phys. Rev. Lett. {\bf 94}, 012005 (2005).

\bibitem{DDmixing}Babar Collaboration, R. Andreassen et al.,
arXiv:0804.0020 [hep-ex]; Belle Collaboration, U. Bitencet al.,
Phys. Rev. {\bf D 77}, 112003 (2008).

\bibitem{Rosner1} J.L. Rosner, hep-ph/0802.1043.

\bibitem{CLEO-f}CLEO Collaboration, Artuso et al.,
Phys. Rev. Lett. {\bf 95}, 251801 (2005); {\bf 99}, 071802 (2007).

\bibitem{Golob} B. Golob (for the Belle Collaboration), Chinese
Phys. {\bf C32} (2008) 495 (proceedings).

\bibitem{CLEO-new} B. Eisenstein et al. The CLEO collaboration,
hep-ex/0806.2112.

\bibitem{LQCD}
 HPQCD Collaboration and UKQCD Collaboration (E. Follana et al.), Phys. Rev. Lett.
  {\bf 100}, 062002 (2008), arXiv:0706.1726 [hep-lat].

\bibitem{He-Deshpande} N.G. Deshpande and X.G. He, Phys. Rev. Lett. {\bf
74}, 26 (1995).

\bibitem{Abbott} L.F. Abbott, P. Sikivie and M.B. Wise, Phys. Rev. {\bf
D 21}, 768 (1980).

\bibitem{w-Dai}W.S. Dai, X.H. Guo, H.Y. Jin and X.Q. Li, Phys. Rev. {\bf D
62}, 114026 (2000).

\bibitem{Li-Yang} H.B. Li and M.Z. Yang, arXiv:0709.0979 [hep-ph].


\bibitem{exp-Ds-pn}CLEO Collaboration, J.P. Alexander et al., Phys. Rev. Lett.
 {\bf 100}, 161804 (2008).

\bibitem{Cheng1}X.Y. Pham, Phys. Lett. {\bf B 94}, 231 (1980);
 Phys. Rev. Lett. {\bf 45}, 1663 (1980);
 I. Bediaga and E. Predazzi, Phys. Lett. {\bf B 275}, 161 (1992).
%H.Y. Cheng, Int. J. Mod. Phys. {\bf A 21}, 4209 (2006).

\bibitem{Cheng2} C.H. Chen, H.Y. Cheng and Y.K. Hsiao, Phys. Lett.{\bf
B 663}, 326 (2008).

\bibitem{y.Wang1} Y. Wang et al. Eur. Phys. J. {\bf C 54}, 107 (2008).

\bibitem{y.Wang2} Y. Wang et al. Eur. Phys. J. {\bf C 55}, 607 (2008).


\bibitem{Stech} M. Bauer, B. Stech and M. Wirbel, Z. Phys. {\bf C
34}, 103 (1989).

\bibitem{Buras} A. Buras, J. Gerard and R. Ruckl, Nucl. Phys. {\bf
B 268}, 16 (1986).

\bibitem{Cheng-1} H.Y. Cheng, Z. Phys. {\bf C 32}, 237 (1986); X.Q. Li, T. Huang
and Z.C. Zhang, Z. Phys. {\bf C 42}, 99 (1989).


\bibitem{Georgi-Carone} H. Georgi and M. Wise, Phys. Lett. {\bf B
243}, 279 (1990); C. Carone, Phys. Lett. {\bf B 253}, 408 (1991).

\bibitem{Guo-Jin-Li} X.H. Guo, H.Y. Jin and X.Q. Li, Phys. Rev. {\bf D 53}, 1153
(1996).

\bibitem{1/mb} Y.B. Dai, X.H. Guo, C.S. Huang and M.Q. Huang, Phys. Rev.
{\bf D 51}, 3532 (1995); T. Mannel, Nucl. Phys. Proc. Suppl. {\bf
39BC} (1995) 426; A. Falk, M. Luke and M.J. Savage, Phys. Rev. {\bf
D 53}, 6316 (1996); C.S. Huang, C. Liu and C.T. Yan, Phys. Rev. {\bf
D 62}, 054019 (2000).

%\bibitem{corrections}
\bibitem{Li-H-n} C. Coriano and H.N. Li, Phys. Lett. {\bf B 309}, 409 (1993);
W.S. Hou, H.N. Li, S. Mishima and M. Nagashima, Phys. Rev. Lett.
{\bf 98}, 131801 (2007); H.N. Li and S. Mishima, Phys. Rev. {\bf D
74}, 094020 (2006).

\bibitem{scet} C.W. Bauer, I.Z. Rothdtein and I.W. Stewart, Phys. Rev. {\bf
D 74}, 034010 (2006); B. Grinstein, Nucl. Phys. {\bf B 775}, 199
(2006); T. Feldmann, arXiv:hep-ph/0610192; F. Liu and J.P. Ma,
arXiv:0802.2973 [hep-ph].

\bibitem{Lu} Y. Wang et al., Eur. Phys. J. {\bf C 54}, 107 (2008); W.
Wang, Y.M. Wang, D.S. Yang and C.D. Lu, arXiv:hep-ph/0805.4695
[hep-ph].

\bibitem{Ma} J.P. Ma and Q. Wang, Phys. Lett. {\bf B 613}, 39 (2005); JHEP
{\bf 0601}, 067 (2006); Phys. Lett. {\bf B 642}, 232 (2006); Phys.
Rev. {D 75}, 014014 (2007).

\bibitem{Ma1}F. Feng, J.P. Ma and Q. Wang, arXiv:0807.0296 [hep-ph].

\bibitem{H-Li-new} H.N. Li, hep-ph/0808.1526.

\bibitem{harmonic}A. Le Yaouanc, L. Oliver, O. P\`{e}ne and J. Raynal,
\textit{Hadron Transitions in the Quark
Model}, Gordon and breach science publishers, New York (1987).

\bibitem{Kanki} T. Kanki, Prog. Theor. Phys. {\bf 56} (1976) 1885; and
many new research progresses have appeared later.


\bibitem{Ebert-feldmann}D. Ebert, T. Feldmann and H. Reinhardt, Phys. Lett. {\bf B 388},
154 (1996); T. Feldmann, arXiv:{hep-ph/9606451}; D. Ebert, T.
Feldmann, R. Friedrich and H. Reinhardt, Nucl. Phys. {\bf B 434}
619 (1995).
\bibitem{Polosa}A.D. Polosa, Riv. Nuovo Cim. {\bf 23N11}, 1 (2000), arXiv:hep-ph/0004183.

\bibitem{HW} C.W. Hwang and Z.T. Wei, J. Phys. {\bf G 34}, 687 (2007).


\bibitem{singlet} L. Clavelli, P.H. Cox and B. Harms, Phys. Rev {\bf
D 29}, 57 (1984); erratum-ibd {\bf 31}, 214 (1985).

\bibitem{evaporation} M. GayDucati, Phys. Lett. {\bf B 464}, 286 (1999).

\bibitem{NRQCD}G.P. Lepage, L. Magnea, C. Nakhleh, U. Magnea and K. Hornbostel,
Phys. Rev. {\bf D 46}, 4052
(1992).

\bibitem{SVZ} M. Shifman, A. Vainstein and V. Zakharov,
Nucl. Phys. {\bf B 147}, 385 (1979).

\bibitem{Shifman} A. Shifman, Proceedings, Workshop, Aachen, Germany, June 9-13, 1992.
Vol. 1, 2.
1993. Singapore, Singapore: World Scientific, 1-869 (1993).

\bibitem{MIT-bag}A. Chodos, R.L. Jaffe, K. Johnson, C.B. Thorn and V.F. Weisskopf,
Phys. Rev. {\bf D 9}, 3471 (1974).

\bibitem{flux-tube} K. Johnson and C. Thorn, Phys. Rev. {\bf D
13}, 1934 (1976); C. Olson, M.G. Olson and K. Williams, Phys. Rev.
{\bf D 45}, 4307 (1992); N. Brambilla and G.M. Prosperi, Phys.
Rev. {\bf D 47}, 2107 (1993).

\bibitem{OPE} X.Q. Li and B.S. Zou, Phys. Lett. {\bf B 399}, 297
(1997).

\bibitem{Dai-YS}Y.S. Dai, D.S. Du, X.Q. Li, Z.T. Wei and B.S.Zou.
Phys. Rev. {\bf D 60}, 014014 (1999).

\bibitem{Cheng-H.Y} L.L. Chau and H.Y. Cheng, Phys. Lett. {\bf B
280}, 281 (1992); S.L. Chen, X.H. Guo, X.Q. Li and G.L. Wang,
Commun. Theor. Phys. {\bf 40}, 563 (2003).

\bibitem{Li-Zou} X.Q. Li and B.S. Zou, Phys. Lett. {\bf B 399}, 297 (1997).

\bibitem{Anisowich}V.V. Anisovich, D.V. Bugg, A.V. Sarantsev and B.S. Zou,
Phys. Rev. {\bf D 51}, R4619 (1995); {\bf D50}, 1972 (1994); M.
Locher and Y. Lu, Z. Phys. {\bf A 351}, 81 (1994); B.S. Zou and
D.V. Bugg, Phys. Rev. {\bf D 48}, R3948 (1993); {\bf D 50}, 591
(1994).



\bibitem{Suzuki} M. Suzuki, Phys. Rev. {\bf D 63}, 054021 (2001).

\bibitem{Liu-Li-Zeng}Xiang Liu, X.Q. Zeng and X.Q. Li, Phys. Rev. {\bf D
74}, 074003 (2006).

\bibitem{Liu-Bc} X. Liu and X.Q. Li, Phys. Rev. {\bf D 77}, 096010 (2008).

\bibitem{OZI} S. Okubo, Phys. Lett. {\bf 5}, 163 (1963); G. Zweig,
CERN Report No. 8182/TH 401; G. Zweig, CERN Report No. 8419/TH-412
(1964); J. Iizuka, Prog. Theor. Phys. Suppl. {\bf 37-8}, 21
(1966).

\bibitem{databook}The Review of Particle Physics, C. Amsler et al., Phys. Lett.
{\bf B 667}, 1 (2008).

\bibitem{Korner} J.G. K\"orner, J.H. Kuhn, M. Krammer and H. Schneider. Nucl. Phys.
{\bf B 229}, 115 (1983); Y.D. Yang, arXiv:hep-ph/0404018.

\bibitem{Tong-Li}G. Li, T. Li, X.Q. Li, W.G. Ma and S.M. Zhao, Nucl. Phys. {\bf B
727}, 301 (2005).

\bibitem{Li-Tong}T. Li, S.M. Zhao and X.Q. Li, arXiv:0705.1195 [hep-ph].

\bibitem{Brodsky}S.J. Brodsky and G.P. Lepage, Phys. Rev. {\bf D 24},
2848 (1981); N. Branbilla et al., arXiv:hep-ph/0412158.

\bibitem{Rosner} J.L. Rosner, Phys. Rev. {\bf D 64}, 094002 (2001).

%\bibitem{Suzuki} M. Suzuki, Phys.Rev. {\bf D 63} (2001) 054021.
\bibitem{Zou-11} X.Q. Li, D.V. Bugg and B.S. Zou, Phys. Rev. {\bf D 55}, 1421 (1997).

\bibitem{X-Mo} X.H. Mo, C.Z. Yuan and P. Wang, arXiv:hep-ph/0611214.

\bibitem{Zhao-Qiang} Q. Zhao, G. Li and C.H. Chang, Phys. Lett. {\bf B 645}, 173 (2007);
G. Li, Q. Zhao and C.H. Chang, J. Phys. {\bf G 35}, 055002 (2008);
Q. Zhao, Talk given at "The conference on interdisciplinary fields
of particle physics, nuclear physics and cosmology", Tenchong,
Yunnan, China, Aug. 2 to 4, will appear at the proceedings of the
conference.

%\bibitem{jpsi-decay} X. Liu and X.Q. Li, Phys.Rev. D
\bibitem{Kuang} Y.P. Kuang, Front. Phys. China, {\bf 1}, 19 (2006).

\bibitem{Yan} T.M. Yan, Phys. Rev. {\bf D 22}, 1652 (1980); Y.P. Kuang and
T.M. Yan, Phys. Rev. {\bf D 24}, 2874 (1981).


\bibitem{CLEOc} CLEO collaboration, J. Phys. Cond. Ser. {\bf 9},
119, (2005).

\bibitem{Li-hc}An incomplete list: X.Q. Li and Xiang Liu, Phys. Rev. {\bf D 74}, 114029 (2006);
S. Godfrey, J. Phys. Conf. Ser. {\bf 9}, 123 (2005).

\bibitem{Fang}An incomplete list: Belle Collaboration, F. Fang et al.,
Phys. Rev. {\bf D 74}, 012007 (2006).

\bibitem{Swanson} E.S. Swanson and A.P. Szczepaniak, Phys. Rev. {\bf D 59}, 014035 (1998).

\bibitem{Allen} T. Allen, M. Olsson and S. Veseli,  Phys. Lett. {\bf B 434},
110 (1998).

\bibitem{chi}C.W. Chiang, M. Gronau, J.L. Rosner and D.A. Suprun,
 Phys. Rev. {\bf D 70}, 034020 (2004).

\bibitem{Close} F.E. Close and A. Kirk, Phys. Lett. {\bf B 483}, 245 (2000);
F.E. Close and Q. Zhao, Phys. Rev. {\bf D 71}, 094022 (2005); X.G.
He, X.Q. Li, Xiang Liu and X.Q. Zeng, Phys. Rev. {\bf D 73},
051502(R) (2006); X.G. He, X.Q. Li, Xiang Liu and X.Q. Zeng, Phys.
Rev. {\bf D 73}, 114026 (2006).

\bibitem{Belle Collaboration}Belle Collaboration, K.F. Chen et al., arXiv:0710.2577 [hep-ex];
K.F. Chen et al., arXiv:0710.2577 [hep-ex].


\bibitem{Meng} C. Meng and K.T. Chao, arXiv:0805.0143 [hep-ph]; arXiv:0712.3595 [hep-ph].

\bibitem{Zou} B.S. Zou, Int. Mod. Phys. {\bf A 21}, 5552 (2006);
 X.G. He, X.Q. Li, Xiang Liu and X.Q. Zeng, Eur. Phys. J. {\bf C 44},
 419 (2005).

\bibitem{Olsen} S. Godfrey and S. Olsen, arXiv:0801.3867 [hep-ph].

%%%%%%%%%%%%%%%%%%%%    2317 %%%%%%%%%%%%%%%%%%%%%%%%%%%%%%%%%%%

\bibitem{2317} The Babar Collaboration, B. Aubert et al., Phys.
Rev. Lett. {\bf 90}, 242001 (2003); F. Porter, Eur. Phys. J. {\bf
C 33}, 219 (2004).

\bibitem{Belle}The Belle Collaboration, P. Krokovny
et al., Phys. Rev. Lett. {\bf 91}, 262002 (2003).

\bibitem{CLEO}The CLEO
Collaboration, D. Besson et al., Phys. Rev. {\bf D 68}, 032002
(2003).

\bibitem{others}Belle Collaboration, Y.Mikami et al., Phys. Rev. Lett. {\bf 92},
012002 (2004); Belle Collaboration, P. Krokovny et al.,
arXiv:hep-ex/0310053; FOCUS Collabortation, E. W. Vaandering,
arXiv:hep-ex/0406044; Babar Collaboration, B. Aubert et al., Phys.
Rev. Lett. {\bf 93}, 181801 (2004); Babar Collaboration, B. Aubert
et al., Phys. Rev. {\bf D 69}, 031101 (2004); Babar Collaboration,
G. Calderini et al., arXiv:hep-ex/0405081; Babar Collaboration, B.
Aubert et al., arXiv:hep-ex/0408067.



\bibitem{Bardeen}W.A. Bardeen, E.J. Eichten and C.T. Hill, Phys.
Rev. {\bf D 68}, 054024 (2003); M.A. Nowak, M. Rho and I. Zahed,
Acta Phys. Polon. {\bf B 35}, 2377 (2004); E. Kolomeitsev and M.
Lutz, Phys. Lett. {\bf B 582}, 39 (2004).

\bibitem{Chaokt}K.T. Chao, Phys. Lett. {\bf B 599}, 43 (2004).

\bibitem{Beverenn1}E. van Beveren and G. Rupp, Phys. Rev. Lett. {\bf 91},
012003 (2003); E. van Bevern and G. Rupp, Eur. Phys. J. {\bf C
32}, 493 (2004).

\bibitem{Narison}S. Narison, Phys. Lett. {\bf B 605}, 319
(2005).
\bibitem{Daiyb}Y.B. Dai, S.L. Zhu and Y.B. Zou, arXiv:{hep-ph/0610327}.

\bibitem{Chen}Y.Q. Chen and X.Q. Li, Phys. Rev. Lett. {\bf93},
232001 (2004); A.P. Szczepaniak, Phys. Lett. {\bf B 567}, 23
(2003); T.E. Browder, S. Pakvasa and A.A. Petrov, Phys. Lett. {\bf
B 578}, 365 (2004).

\bibitem{Cheng}H.Y. Cheng and W.S. Hou, Phys. Lett. {\bf B 566}, 193
(2003).

\bibitem{T. Barnes} T. Barnes, F.E. Close and H.J. Lipkin, Phys. Rev. {\bf D
68}, 054006 (2003); A.P. Szczepaniak, Phys. Lett. {\bf B 567}, 23
(2003).

%\bibitem{huang}M.Q. Huang, Phys. Rev. {\bf D 69}, 114015 (2004).

%%%%%%%%%%%%%%    THEORY

\bibitem{Nielsen}M. Nielsen, arXiv:{hep-ph/0510277}.
\bibitem{zhu}W. Wei, P.Z. Huang and S.L. Zhu, Phys. Rev. {\bf D 73}, 034004 (2006).
\bibitem{decay-1}P. Colangelo and F. De Fazio, Phys. Lett. {\bf
B 570}, 180 (2003).
\bibitem{decay-2}W.A. Bardeen, E.J. Eichten and C.T. Hill, Phys.
Rev. {\bf D 68}, 054024 (2003).
\bibitem{decay-3}S. Godfrey, Phys. Lett. {\bf B 568}, 254 (2003).
\bibitem{decay-4}Fayyazuddin adn Riazuddin, Phys. Rev. {\bf D 69},
114008 (2004).

\bibitem{Liu-2317}X. Liu, Y.M. Yu, S.M. Zhao and X.Q. Li, Eur. Phys. J. C {\bf 47}, 445
(2006).
\bibitem{wang-2317}Z.G. Wang, Phys. Rev. {\bf D 75}, 034013 (2007).

\bibitem{radi-LCQSR}P. Colangelo, F. De Fazio and A. Ozpineci,
Phys. Rev. {\bf D 72}, 074004 (2005).

\bibitem{decay-6}S. Ishida et al., arXiv:{hep-ph/0310061}.

\bibitem{Faessler-2317} A. Faessler, T. Gutsche, V.E. Lyubovitskij and Y.L
Ma, Phys. Rev. {\bf D 76}, 014005 (2007).




%%%%%%%%%%%%%%%%%%%%%%%%%%%%%%%%%   PRODUCTION
\bibitem{huang}M.Q. Huang, Phys. Rev. {\bf D 69}, 114015 (2004).
\bibitem{aliev1}T.M. Aliev and M. Savci, Phys. Rev. {\bf D 73} 114010
(2006).
\bibitem{aliev2}T.M. Aliev, K. Azizi and A. Ozpineci, arXiv:{hep-ph/0608264}.
\bibitem{semi-2317-liu}S.M. Zhao, X. Liu and S.J. Li, Eur.Phy.J. {\bf C 51}, 601
(2007).

\bibitem{liu-psi-2317}X.H. Guo, H.W. Ke, X.Q. Li,
Xiang Liu and S.M. Zhao, Commun. Theor. Phys. {\bf 48}, 509
(2007), arXiv:{hep-ph/0510146}.

%%%%%%%%%%%%%%%%%   2632

\bibitem{selex} The SELEX Collaboration, Phys. Rev. Lett. {\bf 93}
242001 (2004).
\bibitem{0408087}Babar Collaboration, arXiv:{hep-ex/0408087}.
\bibitem{2632-belle}Belle Collaboration, B. Yabsley, arXiv:{hep-ex/0507028}.
\bibitem{Focus}FOCUS Collaboration, R. Kutschke, E831-doc-701-v2,
available at www-focus.fnal.gov.



%%%%%%%%%%%%%%%%%%%%%%%%%%   2860   2715   %%%%%%%%%%

\bibitem{2860-babar}Babar Collaboration, B. Aubert et al.,
 Phys. Rev. Lett. {\bf 97}, 222001 (2006).

\bibitem{belle-2715}Belle Collaboration, K. Abe et al., arXiv:hep-ex/0608031.
%A. Palano, New Spectroscopy with charm quarks
%at B factories, Plenary talk given on 7 June 2006 at the charm
%2006 Intern. Workshop, 5-7 June 2006, Beijing, China.
\bibitem{Belle-2715}Belle Collaboration, J. Brodzicka et al.,
arXiv:0707.3491 [hep-ex].

\bibitem{beveren-2860} E. Van Beveren and G. Rupp, Phys. Rev. Lett. {\bf 97}, 202001 (2006).
\bibitem{Ma-2860}T. Matsuki, T. Morii, K. Sudoh, Eur. Phys. J. A {\bf 31}, 701 (2007).
\bibitem{colangelo-2860}P. Colangelo, F.D. Fazio and S. Nicotri, Phys. Lett.
 {\bf B 642}, 48 (2006).

\bibitem{close-2860}F.E. Close, C.E. Tomas, O. Lakhina and E.S.
Swanson, Phys. Lett. {\bf B 647}, 159 (2007).

\bibitem{potential model}S. Godfrey and N. Isgur, Phys. Rev. {\bf D
32}, 189 (1985).
\bibitem{chiral}M.A. Nowak, M. Rho and I. Zahed, Acta Phys. Polon. B
{\bf 35}, 2377 (2004).

\bibitem{Liu-2860}X. Liu, B. Zhang, W.Z. Deng and S.L. Zhu, Eur. Phys. J. C {\bf 50}, 617
(2007).

\bibitem{Liu-WW-2860}W. Wei, Xiang Liu and S.L. Zhu, Phys. Rev. {\bf D 75}, 014013
(2007).


%%%%%%%%%%%%%%%%%%%%%%%%%%%%%%%%%%%%%%%%%%%%%%

\bibitem{Dubynskiy}S. Dubynskiy, A. Gorsky and M.B. Voloshin, arXiv:0804.2244
[hep-th]; S. Dubynskiy and M.B. Voloshin, arXiv:0803.2224
[hep-ph].


%%%%%%%%%%%%%%%%%%%%%%%%%%%%%%%%%%%%%%%%%%%%%%%%%%%%%%%%%%%%%%%%%%%%%

\bibitem{Bigi-BOOK} I. Bigi and A. Sanda, {\it CP Violation} Cambridge
University Press 2000; arXiv:0808.1773 [hep-ph].

\bibitem{GIM} S. Glashow, J. Illiopolous and L. Maiani, Phys. Rev.
{\bf D 2}, 1285 (1970).

\bibitem{CKM} M. Kabayashi and T. Kaskawa, Prog. Theor. Phys. {\bf
49}, 652 (1973).

\bibitem{Babar-mixing}Babar Collaboration, B. Aubert et al., Pys. Rev. {\bf D 76}, 014018 (2007);
Phys. Rev. Lett. {\bf 98}, 211802 (2007).

\bibitem{Belle-mixing}Belle Collaboration, K. Abe et al., Phys. Rev. Lett. {\bf 98},
 211803 (2007); {\bf 99}, 131803 (2007).

\bibitem{He-mixing}X.G. He and G. Valencia, Phys. Lett. {\bf B
651}, 135 (2007).

\bibitem{Li-mixing}X.Q. Li and Z.T. Wei, Phys. Lett. {\bf B 651}, 380 (2007);
S.L. Chen, X.G. He, X.Q. Li, H.C. Tsai and Z.T. Wei,
arXiv:0710.3663 [hep-ph].

\bibitem{Du}D.S. Du, Eur. Phys. J. {\bf C 50}, 579 (2007).

\bibitem{Xing} Z. Xing, Chinese Phys. {\bf C 32}, (2008) 483
(proceedings).

\bibitem{Yangmz1} H. Li and M. Yang, Phys.Rev. {\bf D 74}, (2006) 094016; X. Cheng et al.
Phys.Rev.{\bf D75}, (2007) 094019.

\bibitem{Georgi:2007ek}H. Georgi, Phys. Rev. Lett.  {\bf 98}, 221601 (2007);
Phys. Lett. B {\bf 650}, 275 (2007).

%%%%%%%%%%%%%%%%%%%%%%%%%%%%%%%%%%%%%%%%%%%%%%%%%%%%%%%%%%%%%%%%%%%%%%%%%%%

%%%%%%%%%%%%%%%   baryons  %%%%%%%%
\bibitem{babar-2880} The BABAR Collaboration, B. Aubert et al., Phys. Rev. Lett. {\bf 98}, 012001
(2007).

\bibitem{belle-2880} The BELLE Collaboration, K. Abe et al., arXiv:
hep-ex/0608043.

\bibitem{babar-2980-3077} The BABAR Collaboration, B. Aubert et al.,
arXiv: hep-ex/0607042.

\bibitem{belle-2980-3077}The BELLE Collaboration, R. Chistov et al.,
Phys. Rev. Lett. {\bf 97}, 162001 (2006).

\bibitem{new-Xi3055}The Babar Collaboration, T. Schr\"{o}der, talk given at the EPS High Energy
Physics Conference, Manchester, July, 2007.

\bibitem{babar-omega}The BABAR Collaboration, B. Aubert et al., arXiv:
hep-ex/0608055.
\bibitem{cleo-2880} The CLEO Collaboration, M. Artuso et al., Phys. Rev.
Lett. {\bf 86}, 4479 (2001).

%%%%%%%%%%%%%%%%%%%%%%%%%%%%%%%%%%  theory
\bibitem{charmed baryons}S. Tawfiq, P.J. O'Donnell, and J.G. K\"{o}rner,
Phys. Rev. {\bf D 58}, 054010
(1998); M.A. Ivanov, J.G. K\"{o}rner, V.E. Lyubovitskij, and A.G.
Rusetsky, Phys. Rev. {\bf D 60}, 094002 (1999); M.Q. Huang, Y.B.
Dai, and C.S. Huang, Phys. Rev. {\bf D 52}, 3986 (1995); ibid.
{\bf D 55}, 7317(E) (1997); S.L. Zhu, Phys. Rev. {\bf D 61},
114019 (2000).
\bibitem{review-charmed}D. Pirjol and T.M. Yan, Phys. Rev. {\bf D 56}, 5483
(1997), and references therein.
\bibitem{review-charmed-1}J.G. K\"{o}rner, D. Pirjol and M. Kraemer, Prog. Part. Nucl.
Phys. {\bf 33}, 787 (1994), and references therein.

\bibitem{charmed-baryons-rosner}J.L. Rosner, arXiv:hep-ph/0612332; arXiv:hep-ph/0609195;
arXiv:hep-ph/0606166.

\bibitem{charmed-baryons-xiang}X.G. He, X.Q. Li, X. Liu and X.Q. Zeng,
arXiv:hep-ph/0606015.

\bibitem{charmed-baryons-cheng}H.Y. Cheng and C.K. Chua, Phys. Rev. {\bf D 75},
014006 (2007).

\bibitem{charmed-baryons-valcarce}H. Garcilazo, J. Vijande and A. Valcarce, arXiv:hep-ph/0703257.

\bibitem{charmed baryons-liu}C. Chen, X.L. Chen, X.Liu, W.Z. Deng
and S.L. Zhu, Phys. Rev. {\bf D 75}, 094017 (2007).
\bibitem{charmed baryons-liu-1} X. Liu, C. Chen, X.L. Chen and W.Z.
Deng, Chin. Phys. {\bf C 32}, 424 (2008), arXiv:0710.0187
[hep-ph].

%%%%%%%%%%%%%%%%%%%%%%%%%%%%%%%%%%%%%%%%%



\bibitem{Bigi}I. Bigi, N. Uraltsev, Phys. Lett. {\bf B 280}, 120 (1992);
I. Bigi, N. Uraltsev and A. Vainshtein, Phys. Lett. {\bf B 293},
430 (1992); (E) {\bf B 297}, 477 (1993); B. Blok and M. Shifman,
Nucl. Phys. {\bf B 399}, 441 (1993); {\bf B 399}, 459 (1993); G.
Belliui et al., Phys. Rep. {\bf 289}, 1 (1997).

\bibitem{Lambdabexp}CDF Collaboration, A. Abulencia et al.,
arXiv:hep-ex/0609021.

\bibitem{Gabbiani} N.G. Uraltsev, Phys. Lett. {\bf B 376}, 303 (1996);
F. Gabbiani, A.I. Onishchenko and A.A. Petrov, Phys. Rev. {\bf D
70}, 094031 (2004); E. Franco, V. Lubicz, F. Mescia and C.
Tarantino, Nucl. Phys. {\bf B 633}, 212 (2002).

\bibitem{Guo-peng}H.H. Shih, S.C. Lee and H.N. Li, Phys. Rev. {\bf D 59}, 094014 (1999);
{\bf D 61}, 114002 (2000); C.H. Chou, H.H. Shih, S.C. Lee and H.N.
Li, Phys. Rev. {\bf D 65}, 074030 (2000); P. Guo, H.W. Ke, X.Q.
Li, C.D. Lu and Y.M. Wang, Phys. Rev. {\bf D 75}, 054017 (2007);
X.G. He, T. Li, X.Q. Li and Y.M. Wang, Phys. Rev. {\bf D 74},
034026 (2006).

\bibitem{SELEX-DOUBLE CHARM}SELEX Collaboration, M. Mattson et al.,
Phys. Rev. Lett. {\bf 89}, 112001 (2002).

\bibitem{SELEX-DOUBLE-1}SELEX Collaboration, J.S. Russ,
arXiv:hep-ex/0209075.
\bibitem{SELEX-DOUBLE-2}SELEX Collaboration, M. Moinester, Czech. J. Phys. {\bf 53} B201
(2003).

\bibitem{SELEX-confirm-double}SELEX Collaboration, A. Ocherashvili,
Phys. Lett. {\bf B 628}, 18 (2005).

\bibitem{SELEX-new CHANNEL}SELEX Collaboration, J. Engelfried,
arXiv:hep-ex/0702001.



\bibitem{lifetime}C.H. Chang, T. Li, X.Q. Li and Y.M. Wang,
arXiv:hep-ph/0704.0016.

%\bibitem{Le Yaouanc}A.Le Yaouanc, L.Olivier, O.P$\grave{e}$ne and
%J.C.Raynal, "Hadron Transitions in the Quark  Model", Gordon and
%Breach Science Publish Publish (1998).

\bibitem{Brambilla} N. Brambilla, A. Vairo and T. Rosch, Phys. Rev.
{\bf D 72}, 034021 (2005).

\bibitem{Kiselev} V. Kiselev and A. Likholded, arXiv:hep-ph/0103169.

\bibitem{double-charm}X. Liu, H.W. Ke, Q.P. Qiao, Z.T. Wei and X.Q. Li,
Phys. Rev. {\bf D 77}, 035014 (2008).


\bibitem{Ma-1} H. Lipkin, Nucl. Phys. Proc. suppl. {\bf 21}, 258 (1991);
H.Y. Gao and B.Q. Ma, Mod. Phys. Lett. {\bf A 14}, 2313 (1999);
S.L. Zhu, Phys. Rev. Lett. {\bf 91}, 232002 (2003); M.L. Yan and
X.H. Meng, Commun. Thor. Phys. {\bf 24}, 435 (1995); and many
other references which are not listed here.

\bibitem{helima}X.G. He, X.Q. Li and J.P. Ma, Phys. Rev. {\bf D 71}, 014031 (2005).


\bibitem{He-pentaquark}X.G. He, T. Li, X.Q. Li and C.C. Lih,
Phys. Rev. {\bf D 71}, 14006 (2005); X.G. He and X.Q. Li, Phys.
Rev. {\bf D 70}, 034030 (2004).

\bibitem{He-mixing-pen} X.G. He, X.Q. Li, X. Liu and X.Q. Zeng,
Eur. Phys. J. {\bf C 44}, 430 (2005).

\bibitem{Jaffe} R. Jaffe and F. Wilczek, Phys. Rev. Lett. {\bf 91}, 232003
(2003); Eur. Phys. J. {\bf C 33}, s38 (2004).

\bibitem{Lipkin} M. Karliner and H. Lipkin, Phys. Lett. {\bf 575}, 249 (2003).

\bibitem{anselmino1} M. Anselmino, P. Kroll and B. Pire, Z. Phys.
{\bf C 36}, 89 (1987); P. Kroll, B. Quadder and W. Schweiger,
Nucl. Phys. {\bf B 316}, 373 (1989); P. Kroll, {\it Proceedings of
the Adriatico Research Conference on Spin and Polarization
Dynamics in Nuclear and Particle Physics}, Triest (1988); P. Kroll
and W. Schweiger, Nucl. Phys. {\bf A 474}, 608 (1987).

\bibitem{Guo} X.H. Guo, A.W. Thomas and A.G. Williams, Phys. Rev. {\bf D
59}, 116007 (1999); {\bf D 61}, 116015 (2000).

\bibitem{Ke} H.W. Ke, et al., in preparation.

\bibitem{Dai} W.S. Dai, X.H Guo. H.Y Jin and X.Q. Li,
Commun. Theor. Phys. {\bf 32}, 127 (1999); {\bf 35}, 50 (2001);
W.S. Dai, X.H. Guo, H.Y. Jin and X.Q. Li, Phys. Rev. {\bf D 62},
114026 (2000).

\bibitem{Falk} A. Falk, M. Luke, M. Savage and M. Wise,
Phys. Rev. {\bf D 49}, 555 (1994).

\bibitem{Ebert} E. Ebert and A.P. Martynenko, Phys. Rev. {\bf D 74}, 054008 (2006).

\bibitem{Lansberg} J. Lansberg, J. Mod. Phys. {\bf A 21}, 3857 (2006).

\bibitem{Lansberg-1} J.~Campbell, F.~Maltoni and F.~Tramontano, Phys.
Rev. Lett. {\bf 98}, 252002 (2007), about QCD corrections to
$J/\psi$ and $Upsilon$ production at hadron colliders;
P.~Artoisenet, J.~P.~Lansberg and F.~Maltoni, Phys. Lett. {\bf B
653}, 60 (2007), about hadronproduction of $J/\psi$ and $\Upsilon$
in association with heavy quark pairs; s-channel cut contribution to
Jpsi production H.~Haberzettl and J.~P.~Lansberg, Phys.\ Rev.\
Lett.\ {\bf 100}, 032006 (2008), about possible solution of $J/\psi$
puzzle.



\bibitem{Chao} Y.J. Zhang and K.T. Chao, Phys. Rev. Lett. {\bf 98}, 092003 (2007).

\bibitem{Wangjx} B. Gong and J.X. Wang, arXiv:0802.3727 [hep-ph]; arXiv:0805.2467 [hep-ph].

\bibitem{Xu-Ye}Y. Xu, PH.D thesis, Instutute of High Energy
Physics, Chinese Academy of Sciences.

\bibitem{Zhangxm} S. Nussinov, R.D. Peccei and X.M. Zhang, Phys. Rev.
{\bf D 63}, 016003 (2000).

\bibitem{Wei-Xu-Li} Z.T. Wei, Y. Xu and X.Q. Li, arXiv:0806.2944 [hep-ph].



\bibitem{Yalsley} B. Yabsley (ed), Proceedings of the charm 2007
workshop, Ithaca, NY, Aug. 5 to 8 (2007).

%\bibitem{Voloshin} M. Voloshin, hep-ph/0711.4556.
\bibitem{Olsen1} S. Olsen, arXiv:0801.1153 [hep-ph].

\bibitem{Valcarce} A. Valcarce, J. Vijiande and N. Barnea, Frascati Physics Series XLVI,
945-952 (2007), arXiv:0711.3114 [hep-ph]; E. Golowich,
arXiv:0806.1868 [hep-ph]; G. Corcella and G. Ferrera,
arXiv:0706.2357 [hep-ph].

\bibitem{Tornqvist} N. T\"{o}rnqvist, Found. Phys. {\bf 11}, 171 (1981).

\bibitem{Bramon} A. Bramon and G. Garbarino, Phys. Rev. Lett. {\bf
88}, 040403 (2002); {\bf 89}, 160401 (2002); J.L. Li and C.F.
Qiao, Phys. Rev. {\bf D 74}, 076003 (2006); Y.B. Ding, J.L. Li and
C.F. Qiao, arXiv:hep-ph/0702271.

\end{thebibliography}
\end{document}